\documentclass[11pt]{article}
\usepackage{aip}
\usepackage{amsmath}
\usepackage{graphicx}
\usepackage{amsfonts}
\usepackage{amssymb}

\begin{document}

\author{Arkady L.Kholodenko\thanks{375 H.L. Hunter Laboratories ,Clemson University,
Clemson ,SC 29634-1905 , USA }}
\title{Random Walks on Figure Eight:From Polymers Through Chaos to Gravity and Beyond}
\date{}
\maketitle
\begin{abstract}
This paper is extended and broadly generalized version of earlier published
rapid communication, Phys.Rev.E \textbf{58}, R5213 (1998).It also elaborates
on some problems which were left unsolved or just mentioned in Phyiscs Reports
\textbf{298}, 251 (1998). Applications of the obtained results include (but
not limited to)polymer physics, classical and quantum chaos, fractional Hall
effect,dynamics of textures in liquid crystals and 2+1 Einsteinian gravity.
The paper presents some self contained excerpts from mathematics (not
discussed so far in physics literature) to facilitate the uninterrupted
reading of the manuscript.

PACS number(s):02.50.-r,05.40.+j,61.41.+e
\end{abstract}

\section{Introduction}

In our recent short paper [1] we had provided new derivation of Spitzer's law
[2] for Brownian windings . This problem is of some importance in polymer
physics since it allows to define the concept of entanglement in precise
mathematical terms. It is also of relewance to other disciplines , e.g. 2+1
gravity , quantum Hall effect, etc. Due to format limitations of our paper[1]
many important topics were either briefly mentioned or not discussed at all.
This paper serves to correct this deficiency. If the reader will have patience
to go over the entire manuscript , he or she will recognize, that, still, this
much longer paper is just an introduction into the very rich field of various
topological problems which are waiting for development. In particular, in this
work we had only briefly mentioned the hyperbolic groups (in section 6) while
this has become a very large area of research in mathematics[3] .Study of
random walks on hyperbolic groups and their associated graphs reveals that,
for example, the notion of ultrametricity which is used in the theory of spin
glasses[4] is just a special case of hyperbolicity.Hence, the static and
dynamic problems associated with spin glasses,glasses in general, and all
other cases which involve replicas, could be studied without replicas by
properly formulating the ultrametric problems within the scope of the
hyperbolic group methods.The steps in this direction were already made by
mathematicians[5] but large amount of work is still waiting for development.
Also, the issues related to symbolic dynamic and finite state automata
interpretation of the results presented in this paper were left completely
outside the scope of this paper. Interested reader can ,in part, correct this
deficiency by reading recent review article by Adler[6].

This paper had been also motivated to some extent by the recent papers in
physics literature where one and the same mathematical object is being treated
by completely different methods. Specifically, in Ref.[7] the authors study
the extension of the results of Aharonov-Bohm [8] to the case of \ two
magnetic fluxes while in Ref.[9] quantum mechanics in the rectangular billiard
with single pointlike scatterer was studied. Subsequently, the results of the
last paper were improved [10] and generalized to many scatterers in Ref.[11].
To this list of papers ,one may add Ref.[12],etc.,while in Ref.[13] the random
walks on the disordered cellular networks were studied. In this paper it is
shown, that all these problems are interrelated and ,therefore, could be
treated using the same mathematical methods.Moreover, other problems could be
studied by these methods as well and some of these problems are discussed in
section 7 . In particular, by analogy with the Aharonov-Bohm problem in two
dimensions which had become a sort of Hydrogen atom-like problem to all
problems which involve physical processes in the multiply connected spaces,in
three dimensions one can think of a problem about random walks in the presence
of nontrivial knots mentioned already in our earlier work, Ref[14]. Such
problem may have at least some biological significance since it is important
to know to what extent diffusion processes are being affected by the topology
of ,say, knotted DNA. In this work, we use the same methods as that developed
for the treatment of random walks on figure eight in order to provide some
solutions to the problem of random walks in the presence of knots. Alternative
methods of solution of this problem are discussed in Ref.[15]. We also
reconsider the dynamics of area-preserving toral homeo(diffeo)morphisms[16]
from the point of view of random walks on the Teichm\"{u}ller modular
group.Such reconsideration is useful for several reasons.First, it allows to
discuss new ways of classifying chaotic and nonchaotic systems through the
associated with them 3-manifolds.Second, it is very useful for the description
of dynamics of textures in liquid crystals and particles in 2+1 gravity . In
this paper,naturally, above topics are studied only in passing so that
interested reader is encouraged to read our recent papers [17] for better
understanding of all aspects of these problems. In this work we demonstrate
how the figure 8 problem is related to the solution of David Hilbert's 21st
problem( in section 5).In the same section we discuss how this problem is
connected with the theory of exactly integrable systems,Knizhnik-Zamolodchikov
equations, quantum groups , theory of elasticity. We would like to emphasize
that our presentation is not just a review of literature:each section contains
worked out examples which are original in nature.At the same time,some
mathematical facts are presented along with these developments to facilitate
the uninterrupted reading of the manuscript.

In section 2 some old results related to random walks on once punctured plane
are revisited from the point of view of the results of Ito and
McKean[18],Lyons and McKean[19] and McKean and Sullivan[20]. The rest of the
paper could be considered as various attempts to extend ,to reanalyze and to
generalize their results to the case of multiple punctures .The guiding
principle in doing this is associated with various aspects of random walks on
groups ,mainly free groups. The comprehensive mathematically rigorous and
physically readable review of \ this subject can be found in Ref.[21]. In our
work only very small portion of the available mathematical results is
utilized. In section 3 we introduce some concepts related to groups and group
presentations \ in order to use them in section 4 devoted to finding the
function which connects the multiply connected trice punctured sphere - the
space in which the random walk on figure 8 is taking place-to the simply
connected universal covering Riemann surface. Although in case of trice
punctured sphere such function can be found ,generalization of the methods by
which it was obtained to more than three punctures represents a formidable
task associated with mathematical methods which so far had been used only in
string theory.We discuss these methods in the same section and later,in
section 7 ,provide an introduction to the completely different approach to the
whole problem based on mathematical ideas of Grotendieck and Belyi [22].The
results of section 4 are being put in a broader context of Riemann-Hilbert
problem and Knizhnik-Zamolodchikov equations in section 5. This section also
clarifies the connections with the gauge field-theoretic treatments of the
same problems popular in physics literature [14].In section 6 we develop some
results of Kasteleyn[23] related to random walks on free groups.Based on this
development, we reobtain the results of Lyons and McKean for
transience/recurrence of random walks on the trice punctured sphere.We also
provide in this section a definition of hyperbolic groups and hyperbolic
spaces and show that the ultrametric spaces(used in the theory of spin
glasses[4]) are just special cases of hyperbolic spaces. The results obtained
in this section are further developed in section 7. In this section, solely
devoted to applications,we discuss several topics. In particular, we discuss
random walks in the presence of knots. We demonstrate that the
transience/recurrence of random walks in the presence of a knot depends upon
its nature.For example, random walk on figure 8 is related to the random walk
in the complement of the figure 8 knot and ,hence,is transient while random
walk in the complement of the trefoil knot(the simplest among torus-type
knots) is recurrent. Next, we discuss the chaotic v.s. regular behavior of
such walks by analyzing their behavior with help of \ number-theoretic
methods, by using the theory of Teichm\"{u}ller spaces and by employing some
facts from the theory of 3-manifolds.These methods provide new efficient ways
of description of chaotic systems and are potentially useful for treatments of
dynamics of liquid crystals and 2+1 gravity. Naturally, these topics are
treated only superficially in this paper, but ,we hope, that many
illustrations which we supply along the way, should be helpful for better
understanding and appreciation of much deeper results available in
mathematics[24] and used in our earlier works[17] for treatment of these
physical problems.This section is finished with several additional
applications. For instance, use of Markov triples , known in the number theory
for more than a century [25] along with the results of Grotendieck and
Belyi[22] provides us with an opportunity to introduce an alternative
explanation of the fractional Hall effect. These results are also
complementary to that presented in section 4. We also discuss some connections
between the results the theory of free groups and the theory of braids . Such
discussion reveals advantages and disadvantages of either methods for studying
evolution of random walks in multiply connected spaces.

\section{Some results about random walks on once punctured plane}

Study of random walks on once punctured plane is closely related to the
Aharonov-Bohm (A-B) effect in quantum mechanics [ 8].Although this problem was
studied very thoroughly, here we would like to emphasize some of its aspects
which, to our knowledge, had received much less attention. In particular,we
would like to investigate how the description of random walks on once
punctured plane $\mathbf{R}^{2}-\mathbf{0}$ is related to the description of
the random walks on the circle $\mathbf{S}^{1}.$The last problem was studied
in great detail by Schulman [26 ,27] and is considered to be the benchmark
problem associated with quantization in multiply connected spaces [28 ].

Although it is intuitively clear that both problems are connected , to
establish such connection mathematically is nontrivial.Let us begin with the
observation that in both cases we are dealing with multiply connected
spaces.The theory of differential equations on the punctured plane was
initiated already by Poincare and is known in the literature as the theory of
Fuchsian equations [29,30 ].According to this theory ,it is essential to find
a simply connected (covering )space which is associated with multiply
connected (base) space.Then, the related problem is solved by lifting it to
this covering (simply connected )space and then,the solution is projected down
to the base space. To find a transformation from the base space space $M$
(multiply connected) to the covering space $\tilde{M}$ (simply connected
)could be already a very difficult (and even unsolvable !) problem
.Accordingly,some simplifications may be required in order to obtain at least
a partial information.The circle $\mathbf{S}^{1}$ and the punctured plane
$\mathbf{R}^{2}-\mathbf{0}$ represent a good examples of use of general
methods outlined above and are ideally suited for study of simplifications.
Fortunately ,both problems could be solved completely so that they could be
considered as having the same status in the theory of diffusion/quantum
mechanical problems in multiply connected spaces as the Hydrogen atom problem
in standard quantum mechanics.

The universal covering space $\tilde{M}$ for the circle problem is real line
\textbf{R }so that \textbf{S}$^{1}=\mathbf{R/Z}$ where \textbf{Z }is the set
of all integers .For the random walk of N effective steps on \textbf{R }we
expect our distribution (Green's) function to be transitionally invariant ,
i.e.
\begin{equation}
G_{N}^{\mathbf{R}}(\tilde{x}^{\prime},\tilde{x})=G_{N}^{\mathbf{R}}(\tilde
{x}^{\prime}-\tilde{x}), \tag{2.1}%
\end{equation}
where $G_{N}^{\mathbf{R}}$ is the Green's function (end-to-end distribution
function) for the random walk on \textbf{R} and $\tilde{x}(\tilde{x}^{\prime
})$ is the initial (final )position of the walk in this space.Surely, Eq.(2.1)
can be equivalently rewritten as
\begin{equation}
G_{N}^{\mathbf{R}}(\tilde{x}^{\prime}+2\pi n,\tilde{x}+2\pi n)=G_{N}%
^{\mathbf{R}}(\tilde{x}^{\prime},\tilde{x}), \tag{2.2}%
\end{equation}
where $n=0,\pm1,\pm2,...$ The above example can be easily generalized now.

Let $\gamma$ be some generator of motion(s) on $\tilde{M}$ ,$\gamma\in\Gamma
,$where $\Gamma$ is some group of motions on $\tilde{M}$ , then,in general, we
should require
\begin{equation}
G_{N}^{\tilde{M}}(\gamma\tilde{x}^{\prime},\gamma\tilde{x})=G_{N}^{\tilde{M}%
}(\tilde{x}^{\prime},\tilde{x}). \tag{2.3}%
\end{equation}
since this result holds for Eq.s(2.1) and(2.2). The above equation can be
equivalently rewritten as
\begin{equation}
G_{N}^{\tilde{M}}(\gamma\tilde{x}^{\prime},\tilde{x})=G_{N}^{\tilde{M}}%
(\tilde{x}^{\prime},\gamma^{-1}\tilde{x}). \tag{2.4}%
\end{equation}
This follows easily if we replace $\tilde{x}$ by $\gamma^{-1}\tilde{x}$ in
Eq.(2.3). In the case of quantum mechanics the transition from $\tilde{M}$ to
$M$ is made via
\begin{equation}
G_{N}^{M}(x^{\prime},x)=\sum\limits_{\gamma\in\Gamma}\rho(\gamma)G_{N}%
^{\tilde{M}}(\gamma\tilde{x}^{\prime},\tilde{x}) \tag{2.5}%
\end{equation}
where $\rho(x)$ is the phase factor ,
\begin{equation}
\rho(\gamma)=e^{i\delta(\gamma)} \tag{2.6}%
\end{equation}
which is subject to the cocycle constraint [31 ,32 ]
\begin{equation}
\delta(\gamma\gamma^{^{\prime}})=\delta(\gamma)+\delta(\gamma^{^{\prime}}).
\tag{2.7}%
\end{equation}
This constraint follows easily from the composition law (Markov property) for propagators.

In the case of punctured plane the phase factor $\delta$ is actually
responsible for changes in statistics and/or for particle-hole interactions.
In the case of \textbf{S}$^{1}$ the phase $\delta=0($or $\delta=\pi)$ is
associated with bosons (or fermions)[26].There is yet another interpretation
of the phase factor.It comes from the polymer physics .In Ref.[14] we had
demonstrated that the description of the random walks on \textbf{S}$^{1}$ can
be associated with the description of random walks (polymers)which are
confined between the parallel plates (or planes)with which these walks are
allowed to interact.The interactions are effectively introducing statistics so
that ,depending on their strength,the polymer chain acts as bosonic or
fermionic quantum particle or as a particle with fractional statistics.

The result given by Eq.(2.5) still deserves some additional comments.Let us
discuss the case of a simple random walk so that we can put $\delta
(\gamma)=0.$Already in Eq.(2.2) we had introduced the fundamental domain $D$:
0$\leq x<2\pi,$ so that the translations producing \textbf{R} are created by
the successive use of $\gamma$ applied to some x$\in D.$This simple result
should hold in more general cases too.That is,we expect that the manifold $M$
should be expressible as a quotient $M=\tilde{M}/\Gamma$ .More intuitively,$M$
should be related tosome fundamental domain in $\tilde{M}$ so that the whole
$\tilde{M}$ is being covered (without gaps!) by translations of the
fundamental domain with help of the group elements $\gamma\in\Gamma.$ Consider
,therefore,instead of Eq.(2.5),its modification
\begin{equation}
\tilde{G}_{N}^{\tilde{M}}(\tilde{x}^{\prime},\tilde{x})=\sum\limits_{\gamma
\in\Gamma}G_{N}^{\tilde{M}}(\gamma\tilde{x}^{\prime},\tilde{x}). \tag{2.8}%
\end{equation}
Let $\gamma^{(1)}$ $\in$ $\Gamma.$ We would like to demonstrate now that
\begin{equation}
\tilde{G}_{N}^{\tilde{M}}(\gamma^{(1)}\tilde{x}^{\prime},\gamma^{(1)}\tilde
{x})=\tilde{G}_{N}^{\tilde{M}}(\tilde{x}^{\prime},\tilde{x})\equiv\tilde
{G}_{N}^{\tilde{M}}(\tilde{x}^{\prime}-\tilde{x}). \tag{2.9}%
\end{equation}
Indeed,using Eq.(2.8),we obtain
\begin{equation}
\tilde{G}_{N}^{\tilde{M}}(\gamma^{(1)}\tilde{x}^{\prime},\gamma^{(1)}\tilde
{x})=\sum\limits_{\gamma\in\Gamma}G_{N}^{\tilde{M}}(\gamma\gamma^{(1)}%
\tilde{x}^{\prime},\gamma^{(1)}\tilde{x}). \tag{2.10}%
\end{equation}
Using Eq.(2.4) ,we obtain as well
\begin{equation}
\sum\limits_{\gamma\in\Gamma}G_{N}^{\tilde{M}}(\left(  \gamma^{(1)}\right)
^{-1}\gamma\gamma^{(1)}\tilde{x}^{\prime},\tilde{x})=\sum\limits_{\gamma
\in\Gamma}G_{N}^{\tilde{M}}(\gamma\tilde{x}^{\prime},\tilde{x}) \tag{2.11}%
\end{equation}
since $\left(  \gamma^{(1)}\right)  ^{-1}\gamma\gamma^{(1)}\in\gamma\in
\Gamma.$This proves Eq.(2.9). Using this result ,it is possible to prove the following

\textbf{Theorem}.\textbf{\ 2.1.}\textit{Let }$G_{N}^{\tilde{M}}(\tilde
{x}^{\prime},\tilde{x})$\textit{\ be the distribution function for the random
walk on the covering space }$\tilde{M}$\textit{\ which possesses property
reflected in Eq.(2.9) ,then the distribution function on the base space }%
$M$\textit{\ can be obtained from }$G_{N}^{\tilde{M}}$\textit{\ through simple
identification: }
\begin{equation}
\tilde{G}_{N}^{\tilde{M}}(\tilde{x}^{\prime},\tilde{x})=G_{N}^{M}(x^{\prime
},x) \tag{2.12}%
\end{equation}
\textit{where }$\tilde{x}^{\prime}$\textit{\ and }$\tilde{x}$ \textit{are
inverse images of }$x$\textit{\ and }$x^{\prime}$\textit{\ under the natural
projection}: $\tilde{M}\rightarrow M=\tilde{M}/\Gamma.$

\textbf{Proof.}\quad Please,consult Ref.[28].\quad\ \ \ \ \ \ \ \ \ \ \ \ \ \ \ \ \ \ \ \ \ \ \ \ \ \ \ \ \ \ \ \ \ \ \ \ \ \ \ \ \ \ \ \ \ \ \ \ \ \ \ \ \ \ \ \ \ \ \ \ \ \ \ \ \ \ \ \ \ \ \ \ \ \ 

\textbf{Remark.} \textbf{2.2}.Using Eq.(2.9),it is clear , that it is
sufficient \ that $\tilde{x}$ and $\tilde{x}^{\prime}\in D.$This means, that
if we are able to find solution for the random walk problem in the covering
space $\tilde{M}$ ,we can obtain solution of the random walk problem in the
base space $M$ without additional complications.

Let us illustrate the above general statements on example of the random walk
on $\mathbf{R}^{2}-\mathbf{0.}$Using dimensionless units we obtain the
following diffusion equation ,
\begin{equation}
\frac{\partial f}{\partial t}=\frac{1}{4}\left(  \frac{\partial^{2}}{\partial
x^{2}}+\frac{\partial^{2}}{\partial y^{2}}\right)  f, \tag{2.13}%
\end{equation}
where the factor $\frac{1}{4}$ (the diffusion coefficient) is chosen for
further convenience and $f=f(x,y;t)$ with $t\rightleftharpoons N.$ It is
convenient to rewrite Eq.(2.13) in terms of complex variables:$z=x+iy,\bar
{z}=x-iy.$ Simple calculation produces then
\begin{equation}
\frac{\partial f}{\partial t}=\frac{\partial^{2}}{\partial z\partial\bar{z}%
}\ f\text{ .} \tag{2.14}%
\end{equation}
Since this equation is not defined for $z=\bar{z}=0$, we would like to
introduce new complex variable $w$ via $z=\exp\{w\}.$ Unlike z,the complex
variable $w=u+iv$ is defined on the entire complex $w-$plane.In terms of
$w-$variable Eq.(2.14) can be rewritten as
\begin{align}
\frac{\partial f}{\partial t}  &  =e^{-2u}\frac{\partial^{2}}{\partial
w\partial\bar{w}}\ f\text{ }\nonumber\\
&  =e^{-2u}\frac{1}{4}\left(  \frac{\partial^{2}}{\partial u^{2}}%
+\frac{\partial^{2}}{\partial v^{2}}\right)  f. \tag{2.15}%
\end{align}
Obtained equation describes Brownian motion on a simply connected covering
space $\tilde{M}$ which is the Riemann surface of the logarithmic
function,i.e.
\begin{equation}
w=\ln\left|  z\right|  +i(\arg z+2\pi n),n=0,\pm1,\pm2,..., \tag{2.16}%
\end{equation}
and is known to be simply connected.The result just obtained coincides with
that discussed in the book by Ito and McKean,e.g.see page 280 of Ref.[18 ].Ito
and McKean argue (without demonstration) that the diffusion Eq.(2.15) can be
converted into the form given by Eq.(2.13) by replacing Brownian time $t$ by
another ,actually, random time $T.$

The idea of the proof goes back to Paul Levi [33] and was refined by others
[34]. Let us discuss in some detail how this can actually be done.To this
purpose , let us rewrite Eq.(2.15) in the following equivalent form
\begin{equation}
\frac{\partial f}{\partial t}e^{2u}=\frac{1}{4}\left(  \frac{\partial^{2}%
}{\partial u^{2}}+\frac{\partial^{2}}{\partial v^{2}}\right)  f. \tag{2.17}%
\end{equation}
Now let us select new time $T$ by requiring \quad%
\[
\frac{dT}{dt}=\exp\{-2u(T)\}
\]
or,equivalently,
\begin{equation}
t=\int\limits_{0}^{T}d\tau\exp\{2u(\tau)\}. \tag{2.18}%
\end{equation}
This then will bring Eq.(2.17) back to the form of Eq.(2.13) as anticipated.
Eq.(2.18) coincides with that presented on page 280 of Ito and McKean [18 ]
where it was given without derivation. In spite of its simple form , thus
transformed diffusion equation will depend upon the random time T since $u(t)$
is Brownian motion. Accordingly,the observables calculated with help of such
diffusion equation will require an extra averaging .This may sometimes be
inconvenient in practical calculations .To recognize this difficulty we shall
discuss a generic example which exhibits all the features involved. Before
doing so ,several remarks are in order.

First, it is clear ,that in more general case of multiple punctures one has to
find an analogue of Eq.(2.16) ,i.e. to find some function $z=\Phi(w)$ such
that $w-$plane (actually Riemann surface) is going to be an universal covering
space $\tilde{M}$ which is simply connected by construction.

Second, to find such function is not an easy task and may not be possible in
general,e.g.sections 4 and 7 below.Nevertheless,if such function can be
found,then, by analogy with Eq.(2.15) ,we will be able to write
\begin{equation}
\frac{\partial f}{\partial t}=\frac{1}{4}\left|  \Phi^{^{\prime}}(w)\right|
^{-2}\left(  \frac{\partial^{2}}{\partial u^{2}}+\frac{\partial^{2}}{\partial
v^{2}}\right)  f \tag{2.19}%
\end{equation}
and,accordingly, the new time can be defined as
\begin{equation}
t=\int\limits_{0}^{T}d\tau\left|  \Phi^{^{\prime}}(w(\tau))\right|  ^{2}
\tag{2.20}%
\end{equation}
so that the diffusion equation ,Eq.(2.19) ,again acquires its canonical form
given by Eq.(2.13) in accord with general result by Paul Levi[33].

Let us now return back to our example of once punctured plane.In this case,
the explicit form of the end-to- end distribution function is well known,e.g.
see Refs.[14,18,35 ] ,and in polar system of coordinates it is given by
\begin{equation}
G(r_{1},r_{2},\Delta\theta;t)=\frac{1}{2\pi t}\exp\{-\frac{r_{1}^{2}+r_{2}%
^{2}}{t}\}\sum\limits_{m=-\infty}^{\infty}e^{im\Delta\theta}I_{m}(z)
\tag{2.21}%
\end{equation}
,where $\Delta\theta=\theta_{2}-\theta_{1},z=2r_{1}r_{2}/t$ and $I_{m}%
(z)=I_{-m}(z)$ is the modified Bessel's function. The above distribution
function can be used either for study of the radial or the angular
distributions or both . Suppose,we are interested in the angular distribution
function only .Then,using Eq.(2.21),it is convenient to introduce the
normalized distribution function defined according to the following
prescription :
\begin{equation}
F(z,\Delta\theta)=\frac{G(r_{1},r_{2},\Delta\theta;t)}{G(r_{1},r_{2}%
,0;t)}=\frac{1}{I_{0}(z)}\sum\limits_{m=-\infty}^{\infty}e^{im\Delta\theta
}I_{\left|  m\right|  }(z). \tag{2.22}%
\end{equation}
The Fourier transform of such defined distribution function can now be
obtained in a standard way as
\begin{equation}
F(z,\alpha)=\int\limits_{-\infty}^{\infty}d\Delta\theta\text{ }e^{-i\alpha
\Delta\theta}F(z,\Delta\theta)=\frac{I_{\left|  \alpha\right|  }(z)}{I_{0}%
(z)}\text{ .} \tag{2.23}%
\end{equation}
Let us now choose $r_{2}=\hat{r}\sqrt{t}+r_{1}.$This choice is motivated by
known scaling properties of Brownian motion [34]. For large $t$ one obtains
:$z\simeq2r_{1}\hat{r}/\sqrt{t}.$ For fixed $\hat{r}$ and $r_{1}$ and
$t\rightarrow\infty$ one surely expects $z\rightarrow0.$Using known asymptotic
expansion of $I_{\left|  \alpha\right|  }(z)$ for small $z^{\prime}s$ in
Eq.(2.23) allows us to obtain closed form analytic expression for
$F(z,\alpha)$ in this limit as
\begin{equation}
F(z,\alpha)\simeq\exp\{-\frac{\left|  \alpha\right|  }{2}\ln t\}. \tag{2.24}%
\end{equation}
The inverse Fourier transform of Eq.(2.24 ) produces famous Cauchy-type
distribution for $\Delta\theta$%
\begin{equation}
P(x=\frac{2\Delta\theta}{\ln t})dx=\frac{1}{\pi}\frac{1}{1+x^{2}}dx \tag{2.25}%
\end{equation}
obtained originally by Spitzer in 1958 [2] in much more complicated way.

Let us now reproduce this result with help of the random time change discussed
above.To this purpose,let us begin with the classical Lagrangian $L$ for the
fictitious Brownian particle traveling in the punctured plane. By introducing
polar system of coordinates our $L$ can be written as
\begin{equation}
L[\tau]=\frac{1}{2}\left(  \dot{r}^{2}+r^{2}\dot{\theta}^{2}\right)  .
\tag{2.26}%
\end{equation}
The path integral for such particle can be written now in a usual way as
\begin{equation}
G(r_{1},r_{2},\Delta\theta;t)=\int D[r[\tau],\theta\lbrack\tau]]\exp
\{-\int\limits_{0}^{t}d\tau L[\tau]\} \tag{2.27}%
\end{equation}
where the limits of path integration are not written explicitly. Following the
same line of arguments as that which had lead us from Eq.(2.13) to (2.15) let
us represent $r-$variable through $x-$variable defined by $r=\exp x.$Surely,
for $-\infty<x<\infty,$we have $0<r<\infty$ as required. The Lagrangian $L$ is
now being converted to
\begin{equation}
L=\frac{1}{2}e^{2x}(\dot{x}^{2}+\dot{\theta}^{2}) \tag{2.28}%
\end{equation}
which,in view of Eq.(2.27) ,should have its analogue in Eq.(2.15).Consider now
the change of time variable inside of the path integral in Eq.(2.27).That is
we have $\tau=\tau(T)$ so that ,for example,
\begin{equation}
\frac{dx}{dT}\frac{dT}{d\tau}\text{ and,accordingly, }d\tau\left(  \frac
{dx}{d\tau}\right)  ^{2}=dT\text{ }\frac{dT}{d\tau}\left(  \frac{dx}%
{dT}\right)  ^{2}. \tag{2.29}%
\end{equation}
This then requires us to make a choice for $\frac{dT}{dt}$ which is identical
with that made in Eq.(2.18) so that in terms of new time $T$ our path
integral,Eq.(2.27), represents, indeed ,a free diffusion process in which $x$
and $\theta$ are two independent Brownian motions in accord with our previous
calculations. Let us now calculate the generating function $F(z,\alpha)$ which
was defined in Eq.(2.23) but ,this time,with help of the propagator Eq.(2.27)
(with time rescaled).We have
\begin{equation}
F(z,\alpha)=\left\langle e^{i\alpha\int\limits_{0}^{T}d\tau\theta(\tau
)}\right\rangle \tag{2.30}%
\end{equation}
where $\left\langle ...\right\rangle $ denotes functional integral averaging
with help of the properly normalized propagator , Eq.(2.27).The Gaussian
integration can be trivially performed now with the result :
\begin{equation}
F(z,\alpha)=\int\limits_{0}^{\infty}dT\int\limits_{x_{1}(0)}^{x_{2}%
(T)}D[x(\tau)]\delta(t-\int\limits_{0}^{T}d\tau\exp\{2x(\tau)\})\exp
\{-\frac{1}{2}\int\limits_{0}^{T}d\tau(\dot{x}^{2}+\alpha^{2})\}. \tag{2.31}%
\end{equation}
In arriving at this result we took into account the constraint ,Eq.(2.18),so
that the final answer is written in terms of the unrescaled time.The
constraints of this sort were discussed before in Refs [36,37].The presence of
constraint complicates calculations somehow.To by- pass this difficulty,the
Laplace transform of Eq.(2.31) can be taken with the result
\begin{equation}
F_{s}(z,\alpha)=\int\limits_{0}^{\infty}dT\int\limits_{x(0)}^{x(T)}%
D[x(\tau)]\exp\{-\frac{1}{2}\int\limits_{0}^{T}d\tau(\dot{x}^{2}+\alpha
^{2}+2s\exp\{2x(\tau)\})\} \tag{2.32}%
\end{equation}
where the Laplace variable s is conjugate to time $\tau$ .The corresponding
Schr\"{o}dinger-like equation can be now easily written as
\begin{equation}
\left[  s\exp\{2x\}-\frac{1}{2}(\frac{\partial^{2}}{\partial x^{2}}-\alpha
^{2})\right]  F_{s}(z,\alpha)=\delta(x-x^{^{\prime}}). \tag{2.33}%
\end{equation}
By using again $r$-variable(instead of $x)$ we obtain instead of Eq.(2.33) the
following equivalent equation ($x\neq x^{\prime})$%
\begin{equation}
\left[  sr^{2}-\frac{1}{2}(r^{2}\frac{\partial^{2}}{\partial r^{2}}%
+r\frac{\partial}{\partial r}-\alpha^{2})\right]  F_{s}(z,\alpha)=0 \tag{2.34}%
\end{equation}
in which one can easily recognize the equation for the modified Bessel's
function(s) $I_{\alpha}$ and $K_{\alpha}$ . The Green's function can be
constructed with help of these functions in a standard way,e.g.see Ref.[18]
,with the result :
\begin{equation}
F_{s}(z,\alpha)=%
\begin{tabular}
[c]{c}%
$I_{\left|  \alpha\right|  }(\sqrt{2s}r_{1})K_{\left|  \alpha\right|  }%
(\sqrt{2s}r_{2})$ \quad\ \ \ \ \ \ \ if\quad$r_{1}<r_{2}$\quad\\
$\ K_{\left|  \alpha\right|  }(\sqrt{2s}r_{1})I_{\left|  \alpha\right|
}(\sqrt{2s}r_{2})$ \quad\quad\quad if $r_{2}<r_{1}$ \ \ \ \ .
\end{tabular}
\tag{2.35}%
\end{equation}
With help of this function the inverse Laplace transform now can be performed
with help of Eq.(56) of Cr.5 of Bateman and Erdelyi [38] resulting in
\begin{equation}
F(z,\alpha)=\frac{1}{2t}\exp\{-\frac{r_{1}^{2}+r_{2}^{2}}{t}\}I_{\left|
\alpha\right|  }(z) \tag{2.36}%
\end{equation}
It is clear ,that this result coincides with earlier obtained Eq.(2.23)
provided that it is properly normalized.

From the above derivation it is clear that the entire distribution
function,Eq.(2.21),can also be obtained. By repeating the same steps as we
have used to arrive at Eq.(2.33) ,we obtain now
\begin{equation}
\left[  s\exp\{2x\}-\frac{1}{2}(\frac{\partial^{2}}{\partial x^{2}}%
+\frac{\partial^{2}}{\partial\theta^{2}})\right]  G(x,x^{^{\prime}}%
,\theta)=\delta(x-x^{^{\prime}})\delta(\theta-\theta^{^{\prime}}). \tag{2.37}%
\end{equation}
This equation is defined in the fundamental domain in the covering space
$\tilde{M}$ which is just strip of finite width 2$\pi$ in $\theta$ direction
and infinite -$\infty<x<\infty$ in x direction. Naturally,this equation
coincides with Eq.(2.17) (if the Laplace transform of Eq.(2.17) is
performed).It is remarkable that the radial and the angular Brownian motions
are completely decoupled in terms of variables which we are using.

If,as before ,we introduce $r$-variable via $r$=$\exp x$ ,then Eq.(2.37)
becomes again Bessel-type equation for the corresponding Green's function.The
fundamental domain D under such change will be converted from an infinite
strip to the whole $r-\theta$ plane with cut along the positive semiaxis.In
this case we observe that Eq.(2.21) is in accord with Eq.(2.12) as
required.Evidently,we could as well use the strip of width $\pi$ ,then the
strip is converted into the upper half plane.The Riemann surface of the
logarithm can be made by properly gluing together either the stuck of cut
$r-\theta$ planes or half planes .That is the Riemann surface can be described
(or constructed) in several ways [38].The half plane description is especially
useful since it allows to obtain the Cauchy-type distribution at once [34].At
the same time,the description in terms of $r-\theta$ variables leads to the
undesirable coupling between the radial and the angular Brownian
motions.Indeed,the Green's function for the circle $S^{1}$of radius $r$ is
known [26,31] to be expressed as
\begin{equation}
G(\Delta\theta,t)=\frac{1}{2\pi}\sum\limits_{m=-\infty}^{\infty}%
e^{im\Delta\theta}e^{-\frac{tm^{2}}{4r^{2}}}\text{ .} \tag{2.38}%
\end{equation}
For $r_{1}=r_{2}$ and $z\gg1$ use of the asymptotic estimate [40]
\[
I_{n}\simeq\frac{1}{\sqrt{2\pi z}}(1+\frac{1}{8z}+O(z^{-2}))\exp
\{z-\frac{n^{2}}{2z}+O(z^{-2})\}
\]
allows us to bring Eq.(2.21) into the form
\begin{equation}
G(r_{1}=r_{2}=r,\Delta\theta;t)\simeq\sqrt{\frac{t}{4\pi r^{2}}}%
\sum\limits_{m=-\infty}^{\infty}e^{im\Delta\theta}e^{-\frac{tm^{2}}{4r^{2}}%
}\text{ .} \tag{2.39}%
\end{equation}
Hence,in terms of usual polar coordinates, it is impossible to disentangle the
angular and the radial parts of Brownian motion for\textbf{\ arbitrary }values
of $r$ and $t$ . At the same time,depending upon the questions being asked,it
may be sufficient to use ,say,the angular part,e.g. Eq.(2.38) only
.Moreover,the obtained results can be generalized to d-dimensions.In this case
the Green's function analogue of $S^{1}$problem is known to be [18]
\begin{equation}
G(\theta_{1,}\theta_{2};t)=\sum\limits_{n=0}^{\infty}e^{-t\gamma_{n}}%
\sum\limits_{l\leq m(n)}S_{n}^{l}(\theta_{1})S_{n}^{l}(\theta_{2}) \tag{2.40}%
\end{equation}
with $\gamma_{n}=\frac{1}{2}n(n+d-2)$ and $S_{n}^{l}(\theta)$ obeying the
generalized spherical harmonics equation
\begin{equation}
\frac{1}{2}\bigtriangledown^{2}S_{n}^{l}=\gamma_{n}S_{n}^{l}\text{ .}
\tag{2.41}%
\end{equation}
The above Green's function should be compared with that for $\mathbf{R}%
^{2}-\mathbf{0}$ .As results of Ref.[18] indicate,the reduction analogous to
that given by Eq.(2.39) is \textbf{not }possible for d%
$>$%
2.

In the case of random walks the most commonly asked questions are associated
with the probability of returning to the origin (or to a given site) and with
the mean time spent at the origin (or at a given site) [41].Unfortunately,the
continuum formulation of the random walk problems is not well suited for
treatment of these problems.Indeed,let us consider the standard
(Gaussian-like) propagator for the random walk in d-dimensions given by [31]
\begin{equation}
G_{N}(\mathbf{x}_{1}-\mathbf{x}_{2})=\left(  \frac{1}{4\pi N}\right)
^{\frac{d}{2}}\exp\{-\frac{(\mathbf{x}_{1}-\mathbf{x}_{2})^{2}}{4N}\}\text{ .}
\tag{2.42}%
\end{equation}
The mean time $<T>$ spent at the origin is defined by the Laplace transform
given by
\begin{equation}
<T>=\lim_{s\longrightarrow0^{+}}\int\limits_{0}^{\infty}dNe^{-sN}%
G_{N}(0)\simeq\int\frac{d^{d}k}{\left(  2\pi\right)  ^{d}}\frac{1}{k^{2}%
}\text{ .} \tag{2.43}%
\end{equation}
Surely,this quantity is not well defined in \textbf{any }dimension.Use of
spherical coordinates converts Eq.(2.43) into
\begin{equation}
<T>\approx\int\limits_{0}^{\infty}dkk^{d-3}\equiv G(0)\text{ .} \tag{2.44}%
\end{equation}
More accurate lattice calculations indicate [42] that the above integral is
actually finite for d=3 and divergent for d=1 and 2.The probability $\Pi(0)$
of returning to the origin is known to be related to $G(0)$ as follows [41,42]
:
\begin{equation}
\Pi(0)=1-\frac{1}{G(0)}\text{ .} \tag{2.45}%
\end{equation}
accordingly,the random walk is \textbf{recurrent} or \textbf{transient}
depending upon $\Pi(0)$ being equal or lesser than one.

Going back to our initial problem we would like to find out if the recurrence
(or transience ) for $S^{1}$ problem can help us to establish the
recurrence/transience for \textbf{R}$^{2}-\mathbf{0}$ problem. Looking at Eqs.
(2.38),(2.39) we notice that, at least for z$\gg1,$ such questions are
legitimate to ask.

Using Eq.(2.38) for $\Delta\theta=0$ we obtain,
\begin{equation}
G(0)=\frac{2r^{2}}{\pi}\sum\limits_{m=-\infty}^{\infty}\frac{1}{m^{2}}%
=\infty\text{ .} \tag{2.46}%
\end{equation}
From here we conclude that the random walk on $S^{1}$is recurrent.Let us
consider now the full expression, Eq.(2.21) ,for the propagator in
\textbf{R}$^{2}-\mathbf{0}$ plane.We obtain,
\begin{equation}
G(0)=\frac{1}{2\pi}\sum\limits_{m=-\infty}^{\infty}\int\limits_{0}^{\infty
}\frac{dt}{t}\exp\{-\frac{r^{2}}{t}\}I_{m}(\frac{2r^{2}}{t})\text{ .}
\tag{2.47}%
\end{equation}
For integer m we get $I_{m}(z)=I_{-m}(z).$ Therefore ,to evaluate $G(0)$ we
can use the series expansion for $I_{m}(z),m\geq0:$%
\begin{equation}
I_{m}(z)=\sum\limits_{n=0}^{\infty}\frac{\left(  \frac{z}{2}\right)  ^{2n+m}%
}{n!\Gamma(m+n+1)}\text{ .} \tag{2.48}%
\end{equation}
Substitution of this expression into Eq.(2.47) and use of the standard tables
of integrals provides us with generic result:
\begin{align*}
I  &  =\int\limits_{0}^{\infty}dxx^{\nu-1}\exp\{-\frac{\beta}{x}-\gamma x\}\\
&  =2\left(  \frac{\beta}{\gamma}\right)  ^{\frac{\nu}{2}}K_{\nu}(2\sqrt
{\beta\gamma})
\end{align*}
for the integral in Eq.(2.47) which ,when combined with Eqs.(2.47) and (2.48)
leads us to the same conclusion :$G(0)=\infty$ .Using different
arguments,McKean and Lyons [19] have reached the same conclusion: Brownian
motion on once punctured plane is recurrent.This means,in particular,that the
polymer lying in the plane \textbf{cannot} be entangled with another polymer
which is perpendicular to this plane.The planarity of the problem is not too
restrictive ,e.g. see Ref.[14].This means that the same result will remain
correct in case of cylindrically symmetric problem.The situation changes
drastically if,instead of \textbf{one} hole ,we would consider planar Brownian
motion in the presence of \textbf{two} holes.This problem and its
generalizations is considered in the rest of this paper.

\section{ The trice punctured sphere as hyperbolic Riemann surface}

In the previous section the task of finding the universal covering surface
$\tilde{M}$ for the once punctured plane was solved rather
straightforwardly.The mapping $z=\exp w$ \quad(or $w=\ln\left|  z\right|
+i(\arg z+2\pi n))$ with $z\in M$ and $w\in\tilde{M}$ provides the desired
answer.This answer was obtained,however,not in a systematic way but rather by
a simple guessing.If two (or more) punctures are involved ,a simple guessing
can hardly help.Therefore,to find the covering space $\tilde{M}$ and the
mapping function $z=f(w)$ becomes a problem on its own.There are many ways to
arrive at correct answer (which for the case of arbitrary number and
arrangement of punctures may not even exist ,e.g.see Section 4).In this work
we choose the group-theoretic method which is directly associated with the
homotopy theory used for treatment of diffusion on the circle $S^{1}$
discussed in Section 2.An excellent introduction to the homotopy (and also to
the knot) theory is Ref.[43] to which we refer our readers for more
comprehensive treatment(s).Some facts about the combinatorial group theory can
be found in accessible form in Refs[44,45].

Let us begin with the following observation.The once punctured sphere
$S^{2}-\mathbf{0}$ is actually simply connected(i.e.each point can be
connected with any other so that all paths are homotopically equivalent).The
twice punctured sphere is equivalent to \textbf{R}$^{2}-\mathbf{0}$ and
,therefore,the covering problem we had studied already in previous section.The
homotopy classes in this case are made of paths with different winding
numbers.The homotopy group is free infinite cyclic and abelian with just one
generator $a$ so that paths with different winding numbers are made of
successive applications of $a$ and $a^{-1}.$Surely, $a^{m}a^{n}=a^{n}a^{m}$
for windings $m$ and $n$ since the group is abelian.Let us now consider the
trice punctured sphere (or twice punctured plane).The homotopically
nonequivalent paths are depicted in Fig.1.%

\begin{figure}
[ptb]
\begin{center}
\includegraphics[
height=2.2866in,
width=2.3661in
]%
{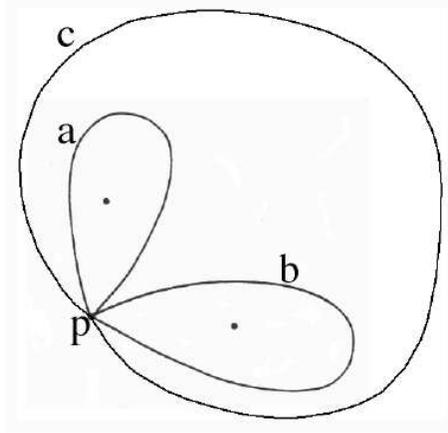}%
\caption{Homotopy of paths(with respect to some arbitrary point p) on the
trice punctured spere}%
\end{center}
\end{figure}
The homotopy group is made of generators $a$ and $b$ related to the windings
around the first and the second puncture respectively. Lyons and McKean [19]
and McKean and Sullivan [20] in their study of diffusion on the trice
punctured sphere introduce yet another generator $c$ which is associated with
windings around the point at infinity. The homotopy group is non-abelian with
3 generators $a,b$ and $c$ subject to one relation
\begin{equation}
abc=1 \tag{3.1}%
\end{equation}
The combinatorial group theory teaches us [44] ,that\textbf{\ any }group $G$
can be described in terms of set of generators $\{a_{i}\}$ and relations
$\{R_{i}\}$ between some (or all) of group elements.The generators and
relations form a \textbf{presentation} for the group $G$ ,i.e.
\begin{equation}
G=<a_{1},...a_{n}\mid R_{1},...,R_{k}>. \tag{3.2}%
\end{equation}
In the cases when there are no relations the groups are called \textbf{free.}
It can be proven that \textbf{any }free group with two or more generators is
non-abelian. It also can be proven that any group $G$ can be considered as
some subgroup of the free group (with sufficient number of generators).In our
case we have
\begin{equation}
G=<a,b,c\mid abc=1> \tag{3.3}%
\end{equation}
This is a special case of the so called triangle groups [45] to be further
discussed in section 7. At first sight it looks like this group
is\textbf{\ not} free.This ,however, is \textbf{not }the case. Indeed, let us
notice first that $c$=$\left(  ab\right)  ^{-1}$ .Consider now the
combinations $ab$ and $ba$ .If the relation $abc=1$ would be absent
and,instead ,we would have just free group made of two elements $a$ and $b$
,then we can construct a word $W$ (actually ,the \textbf{reduced} word) via
\begin{equation}
W=a^{\alpha_{1}}b^{\beta_{1}}...a^{\alpha_{r}}b^{\beta_{r}} \tag{3.4}%
\end{equation}
where $\alpha_{1}$ and $\beta_{r}$ can be \textbf{any} integers while
$\alpha_{i\text{ }}$ and $\beta_{i\text{ }}$ are any integers ,\textbf{except
zero}.Evidently,the word $W$ encodes some specific pattern of windings around
the first and the second puncture.If,instead of elements $a$ and $b$ we would
consider the combinations $\hat{a}=ab$ and $\hat{b}=ba$ ,then,we can again
construct some word,just like in Eq.(3.4).The totality of all words based on
generators $\hat{a}$ and $\hat{b}$ happens to be the same as that based on
generators $a$ and $b$ [44]. But $c$ is just $\hat{a}^{-1}$! Because of
this,the group based on presentation given by Eq.(3.3) is actually free group
constructed of two generators, i.e.presentation given by Eq.(3.3) can be
replaced by the equivalent presentation
\begin{equation}
G=<a,b\mid>. \tag{3.5}%
\end{equation}
It can be shown[44], that such defined free non-abelian group can be realized
with help of ,say, 2$\times2$ matrices
\begin{equation}
a\Longrightarrow\left\|
\begin{tabular}
[c]{cc}%
1 & 0\\
x & 1
\end{tabular}
\right\|  \text{ \quad and \quad}b\Longrightarrow\left\|
\begin{tabular}
[c]{cc}%
1 & x\\
0 & 1
\end{tabular}
\right\|  \tag{3.6}%
\end{equation}
so that
\begin{equation}
a^{n}=\left\|
\begin{tabular}
[c]{cc}%
1 & 0\\
nx & 1
\end{tabular}
\right\|  \text{ \quad and \quad}b^{n}=\left\|
\begin{tabular}
[c]{cc}%
1 & nx\\
0 & 1
\end{tabular}
\right\|  \tag{3.7}%
\end{equation}
where x is some rational number.The above introduced matrices are
unimodular.This means the following. Let z be some complex number.The group of
unimodular transformations is made of transformations of the type
\begin{equation}
z\rightarrow\frac{az+b}{cz+d} \tag{3.8}%
\end{equation}
such that the determinant $DetL=ab-bc$ of the matrix
\begin{equation}
L=\left\|
\begin{tabular}
[c]{cc}%
a & b\\
c & d
\end{tabular}
\right\|  \tag{3.9}%
\end{equation}
is equal to one.Clearly,transformations defined by matrices $a$ and $b$ are
unimodular. It is known [46] that the unimodular transformations are acting on
the upper half plane( Poincare model $H^{2}$of the hyperbolic plane )defined
by
\begin{equation}
H^{2}=\{x+iy=z\in C\mid y>0\} \tag{3.10}%
\end{equation}
by mapping it to itself.

The $H^{2}$ is the universal covering space for every Riemann surface of genus
$g>1[46,47].$ Such surface can be constructed with help of some fundamental
polygon $D$ in $H^{2}.$The appropriately chosen(see below) unimodular
transformations will translate(tesselate) this polygon across $H^{2}$ so that
the entire $H^{2}$ will be covered by copies of this polygon \textbf{without
gaps. }This picture is an elementary extension of a much simpler picture of
torus as a quotient of the Euclidean plane by the group of translations
tesselating the fundamental quadrangle or of a circle $S^{1}$as quotient
\textbf{R/Z} which we had studied in Section 2.

For the trice punctured sphere we can choose one puncture at infinity while
two others at points 0 and 1 respectively.Then,the fundamental triangle is
depicted in Fig.2.%

\begin{figure}
[ptb]
\begin{center}
\includegraphics[
height=2.437in,
width=3.2949in
]%
{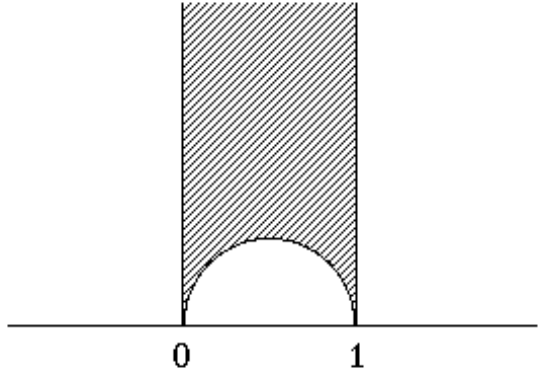}%
\caption{The fundamental triangle on H$^{2}$}%
\end{center}
\end{figure}
Naturally,the vertices of this triangle should not belong to the triangle
itself. If the $H^{2}$plane is filled by such tesselated triangles,the
$w$-plane is covered by an infinity of the upper and lower half-planes which
are the conformal images of the tesselated triangles.Each half-plane has three
''neighbors'' which are connected with it along the stretch $0<w<1$ and two
rays :-$\infty<w<0$ and 1$<w<\infty$ respectively.The totality of half-planes
connected with each other in the way just indicated forms the modular surface
$\tilde{M}$ (one way of making Riemann surfaces[48]). The modular surface
$\tilde{M}$ is the universal covering surface for the trice punctured sphere.
More exactly, the motion of some Brownian walker on the twice punctured plane
around puncture(s) can be associated with the motion of its image on $H^{2}$
.The topology of the twice punctured plane model is actually the same as for
the once punctured torus .This fact was briefly discussed in our earlier work
, Refs[1,14] .

Let us discuss now some consequences of this observation.In the next section
we are going to find the explicit form of the function which maps the complex
plane with cuts (or better $\tilde{M}$ )into $H^{2}$ . Let $w=J(z)$ be such
function so that $w\in\tilde{M}$ and $z\in H^{2}$ .Consider action of such
defined function on the fundamental triangle depicted in Fig.2.without loss of
generality ,it is more convenient to consider two adjacent triangles as
depicted in Fig.3.%

\begin{figure}
[ptb]
\begin{center}
\includegraphics[
height=2.4889in,
width=2.8867in
]%
{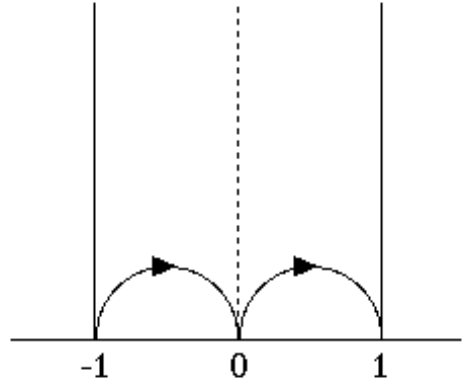}%
\caption{The fundamental quadrangle on H$^{2}$ .It is also used for
description of the punctured torus,e.g.see Ref.[] and Fig.13 below.}%
\end{center}
\end{figure}
These triangles make up the region R$\subset H^{2}$ bounded by two vertical
lines Rez=$\pm1$ and two semicircles $\left|  z\pm\frac{1}{2}\right|
=\frac{1}{2}.$ The left boundary Re$z=-1$ is connected with the right boundary
Re$z=1$ with help of translation: $b(z)=2+z$ while the lower left boundary
$\left|  z+\frac{1}{2}\right|  =\frac{1}{2}$ is connected with the lower right
$\left|  z-\frac{1}{2}\right|  =\frac{1}{2}$ by means of transformation
$a(z)=\frac{z}{2z+1}$ .Comparison with Eqs(3.6) and (3.8) indicates that the
combination of such defined motions do constitute free non-abelian group
composed of two generators. For the function $J(z)$ we have to require
\begin{equation}
J(a(z))=J(z) \tag{3.11a}%
\end{equation}
and also
\begin{equation}
J(b(z))=J(z) \tag{3.11b}%
\end{equation}
in order to achieve the tesselation of $H^{2\text{ }}$without gaps.It could be
shown as well that
\begin{equation}
J(ab(z))=J(ba(z))=J(z) \tag{3.12}%
\end{equation}
Such equations are the defining relations for the automorphic function $J(z).$
At this point,several questions arise. First,are the above transformations
$a(z)$ and $b(z)$ unique ?Second,based on these transformations alone can one
find the explicit form of the function $J(z)$ by using the periodicity
properties shown above ? The answer to the first question is negative.Already
Eqs.(3.6) and (3.7) indicate that this is \textbf{not }the case.At the same
time, for given set of $a$ and $b$ generators it is possible to find the
automorphic function $J(z)$ which will possess the required transformation
properties. Non uniqueness of the transformations leads to the non uniqueness
of the fundamental polygon $D$ on which these generators act.Recent studies
had shown [49] that some choice of generators of the free group G($\mu)$,e.g.
\begin{equation}%
\begin{tabular}
[c]{cc}%
$b(z)=2+z$ & \\
$a(z)=\mu+\frac{1}{z}$ &
\end{tabular}
\tag{3.13}%
\end{equation}
where $\mu_{1}=3i;$or $\mu_{2}=0.0533+1.9i$ $,$etc also describe the homotopy
group of the punctured torus. For the case $\mu_{1}=3i$ a simple connected
domain D$\in H^{2}$ is depicted in Fig.4.%

\begin{figure}
[ptb]
\begin{center}
\includegraphics[
height=2.2866in,
width=3.8363in
]%
{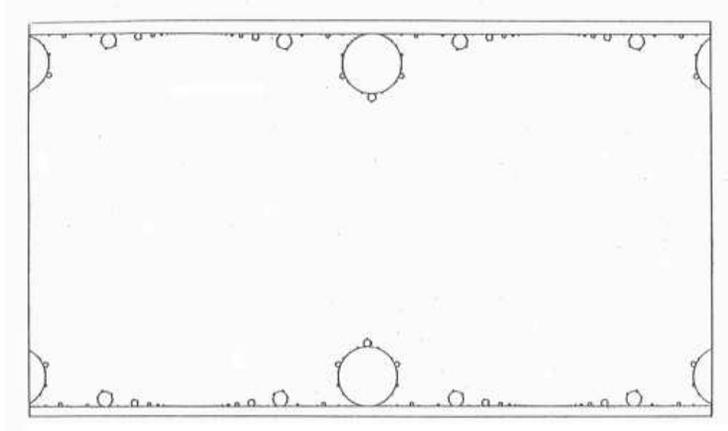}%
\caption{A simply connected domain D of H$^{2}$ whose quotient D/G($\mu_{1}$)
produces the punctured torus []}%
\end{center}
\end{figure}
Action of generators $a(z)$ and $b(z)$ on the fundamental polygon \textbf{does
not} lead to the tesselation of the entire $H^{2}$ .Instead ,the above free
group tesselates some subdomain $\Delta$ of $H^{2}$ so that the quotient
$\Delta/G$ is still the punctured torus.For $\mu=\mu_{2}$ the domain $\Delta$
is depicted in Fig.5
\begin{figure}
[ptbptb]
\begin{center}
\includegraphics[
height=2.2866in,
width=4.1182in
]%
{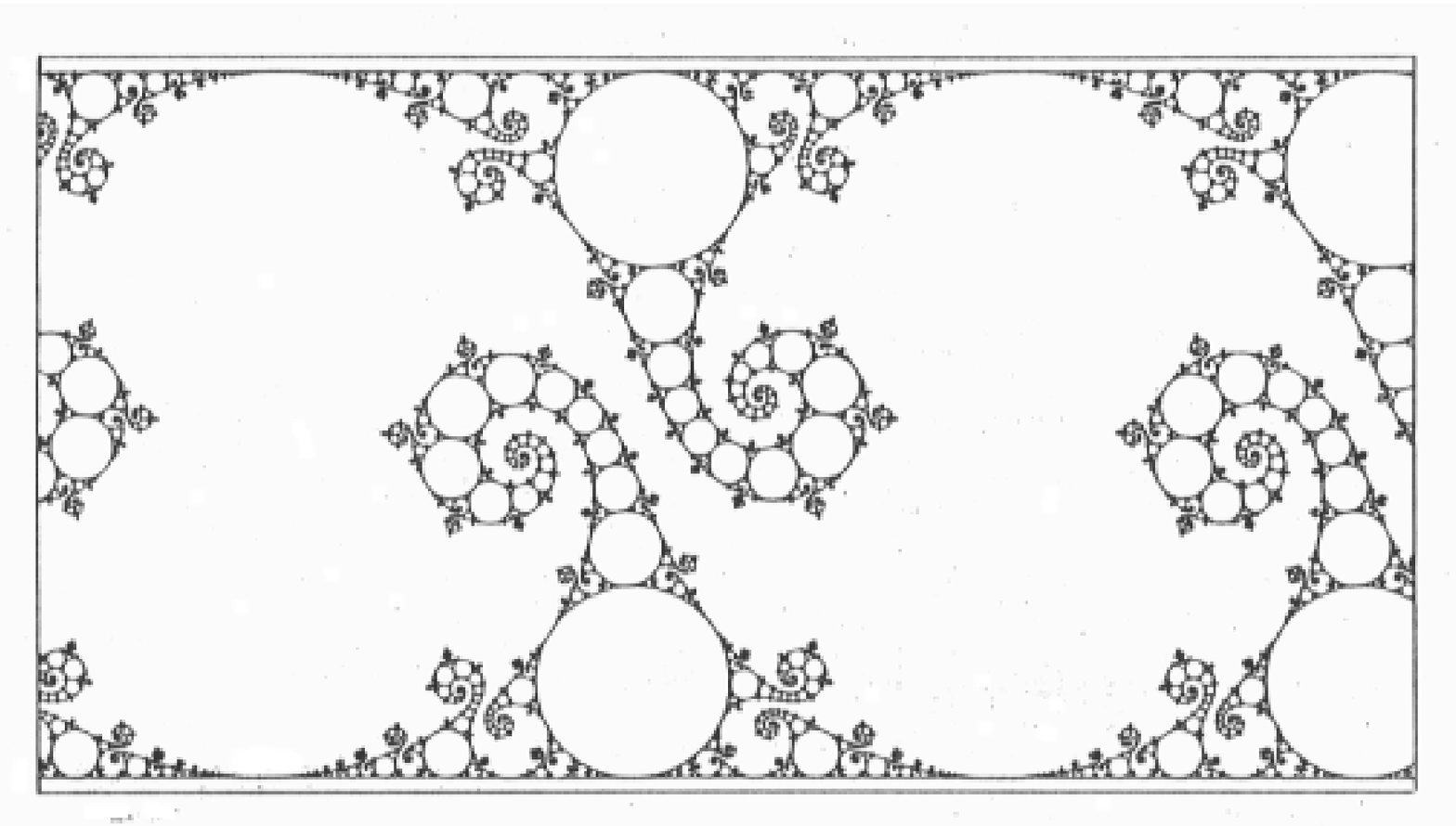}%
\caption{A simple connected domain D of H$^{2}$ whose quotient D/G($\mu_{2})$
produces puncured torus as well.[].}%
\end{center}
\end{figure}
and has, by now familiar,fractal-like shape.We shall not go into details of
such much more elaborate treatment of the punctured torus .For less exotic
situations ,e.g.that depicted in Fig.3,the explicit form of the function
$J(z)$ can be found and this task is accomplished in the next section.

\section{Uniformization of the n-punctured sphere}

On the sphere $S^{2}=C\cup\{\infty\}$ let us consider a region $\Omega$
obtained by removing a finite number of points, i.e.$\Omega=S^{2}%
\backslash\{p_{1},...,p_{n}\}$ with n$\geq3$ .According to the uniformization
theorem there must be some meromorphic function $\lambda$ which provides the
universal covering map of $\Omega$ by $H^{2}$ defined by Eq.(3.10). Suppose
$J$ is some function which maps $H^{2}$ into $\Omega$ ,then if $g$ is some
function which maps $H^{2}$ to $H^{2}$ (e.g.see Eq.(3.8)), then $J=\lambda
\circ g$ ,where $\circ$ denotes the functional composition. Surely, such
defined function $J$ coincides with that which was introduced earlier,e.g.see
Eq.s (3.11),(3.12).The function $J$ is locally invertible so that a closed
path in $\Omega$ has as its image a closed path in $H^{2}$ if and only if it
is homotopic to zero.A closed path in $\Omega$ with ,say ,winding number +1
with respect to $p_{k}$ and 0 with respect to $p_{j}$, $j\neq k$ ,lifts to a
path in $H^{2}$ with the point $T_{k}$ $z$ ,where $T_{k\text{ }}$ is related
to one of the homotopy generators ,e.g. $a$ (or $b$ ) , discussed in the
previous section. Evidently,by analogy with Eq.(3.1), we can write in this
more general case
\begin{equation}
T_{1}T_{2}...T_{n}=1 \tag{4.1}%
\end{equation}
Let us now connect points $p_{1}$and $p_{2}$ by a simple (without crossings)
arc $q_{1}$ and let us proceed in the same fashion with $p_{2}$ and
$p_{3\text{ }},...,p_{n}$ and $p_{1}.$ Then,$\Omega$ is being decomposed into
two simply connected regions $G_{1}$ and $G_{2}.$ Upon mapping into $H^{2}$
these regions will go into subregions of $H^{2}$ whose boundary meets the real
line ($y=0)$ at 2n-2 points.Fig.3 represents just an example of this sort.In
this case we have 3 points -1,0 and +1 of the fundamental region $R$ located
on the real line while the fourth point is located at $\pm\infty$ of real
line.In order to find such mapping the following theorem is helpful[50].

\textbf{Theorem.4.1.} \textit{Suppose }$z=J(w)$\textit{\ where }$w\in H^{2}%
$\textit{\ is universal covering map of }$\Omega,$\textit{then,\{w,z\}is
meromorphic function of z given by}
\begin{equation}
\{w,z\}=\frac{1}{2}\sum\limits_{k=1}^{n}\frac{1}{\left(  z-p_{k}\right)  ^{2}%
}+\sum\limits_{k=1}^{n}\frac{m_{k}}{z-p_{k}} \tag{4.2}%
\end{equation}
\textit{where the Schwarzian derivative }$\{w,z\}$\textit{\ is defined by}
\begin{equation}
\{w,z\}=\left[  \left(  \frac{w^{^{\prime\prime}}}{w^{\prime}}\right)
^{^{\prime}}-\frac{1}{2}\left(  \frac{w^{^{\prime\prime}}}{w^{\prime}}\right)
^{2}\right]  \tag{4.3}%
\end{equation}
\textit{with }$w=w(z)\equiv J^{-1}(z)$\textit{\ and }$w^{\prime}=\frac{dw}%
{dz},$\textit{etc ,and constants }$m_{k}$\textit{\ are subject to relations:}
\[
\sum\limits_{k=1}^{n}m_{k}=0,
\]%
\begin{equation}
\sum\limits_{k=1}^{n}(2m_{k}p_{k}+1)=0, \tag{4.4}%
\end{equation}%
\[
\sum\limits_{k=1}^{n}(m_{k}p_{k}^{2}+p_{k})=0.
\]
Eq.4.2 is the third order nonlinear differential equation whose solution may
or may not be easy to find. However,already Poincar$e^{\prime}$ had discovered
[29] that ,instead of struggling with solution of Eq.(4.2) ,one can consider
much simpler second order linear differential equation of the type
\begin{equation}
y^{\prime\prime}+\frac{1}{2}\{w,z\}y=0. \tag{4.5}%
\end{equation}
In order to solve Eq.(4.5) the r.h.s.of Eq.(4.2) must be used.Then,the
obtained equation is of Fuchsian type[30] and can be solved,provided that the
constants $m_{k}$ (the accessory parameters) are known. Eq.(4.5) produces two
linearly independent solutions $y_{1}$ and $y_{2}.$The most spectacular
outcome of all this can be summarized as follows:

\textbf{Theorem} \textbf{4.2} . \textit{If two independent solutions of
Eq.(4.5) are known, then the mapping function }$w=w(z)$\textit{\ can be
determined from solutions of one of the following} \textit{equations}:
\begin{equation}
y_{1}=w\left(  w^{\prime}\right)  ^{-\frac{1}{2}}\text{ \quad or \quad}%
y_{2}=\left(  w^{\prime}\right)  ^{-\frac{1}{2}}. \tag{4.6}%
\end{equation}

To find the accessory parameters $m_{k}$ is not an easy task in general.
Recent results of conformal and string theory provide some new results in this
direction. Before discussing these results,let us illustrate how the results
just described work in the case of twice and trice punctured sphere. In the
first case we may associate $p_{1}$ with $z=0$ so that the combined use of
Eqs(4.2)-(4,5) produces the following differential equation:
\begin{equation}
y^{\prime\prime}+\frac{1}{4z^{2}}y=0\text{ .} \tag{4.7}%
\end{equation}
Looking for solution of Eq.(4.7) in the form $y=z^{\alpha}$ we can easily
obtain $y_{1,2}=\sqrt{z}$ .Using Eq.(4.6) we obtain as well
\begin{equation}
\sqrt{z}=\left(  \sqrt{w^{\prime}(z)}\right)  ^{-1} \tag{4.8}%
\end{equation}
This leads us immediately to
\begin{equation}
w=\ln\left|  z\right|  +const \tag{4.9}%
\end{equation}
The obtained result is incomplete agreement with earlier obtained,Eq.(2.16).

In the second case we may associate $p_{1}$ with $z=0$, $p_{2}$ with $z=1$ and
$p_{3}$ with $\ z=\infty.$ The combined use of Eqs(4.2)-(4.5) then produces
\begin{equation}
y^{\prime\prime}+\{\frac{1}{z^{2}}+\frac{1}{4(z-1)^{2}}+\frac{1}{4z}-\frac
{1}{4\left(  z-1\right)  }\}y=0. \tag{4.10}%
\end{equation}
This equation can be reduced to that for the hypergeometric function if we use
the following substitution:
\begin{equation}
y(z)=\sqrt{z(z-1)}f(z). \tag{4.11}%
\end{equation}
Then, for the function $f(z)$ we obtain the following hypergeometric
equation:
\begin{equation}
z(z-1)f^{\prime\prime}+(1-2z)f^{\prime}-\frac{1}{4}f=0. \tag{4.12}%
\end{equation}
Let us recall that the hypergeometric function $F$ obeys in general equation
of the type
\begin{equation}
z(z-1)F^{\prime\prime}+[c-(a+b+1)z]F^{\prime}-abF=0 \tag{4.13}%
\end{equation}
which has solution in terms of the following series expansion
\begin{equation}
F(a,b,c;z)=1+\frac{a\cdot b}{c\cdot1}z+\frac{a\cdot(a+1)\cdot b\cdot
(b+1)}{c\cdot(c+1)\cdot1\cdot2}z^{2}+... \tag{4.14}%
\end{equation}
For a special values of $a,b$ and $c$ and also $z$ this expansion coincides
exactly with the series expansion for the complete elliptic integral $K$
defined by
\begin{equation}
K=\int\limits_{0}^{\frac{\pi}{2}}dx\frac{1}{\sqrt{1-k^{2}\sin^{2}x}}.
\tag{4.15}%
\end{equation}
For this to happen,we have to require $a=b=\frac{1}{2},c=1$ and $z=k^{2}$ in
Eq.(4.13).This then produces [48]
\begin{equation}
K=\frac{\pi}{2}F(\frac{1}{2},\frac{1}{2},1;k^{2}) \tag{4.16}%
\end{equation}
Using these results we conclude that $K$ is solution of Eq.(4.12) (since the
values of constants $a,b$ and $c$ produce Eq.(4.12)) in which we have to
replace $z$ by $k^{2}.$ Another independent solution of Eq.(4.12) can be
obtained if we replace $k^{2}$ in Eq.(4.16) by $\left(  k^{\prime}\right)
^{2}=1-k^{2}.$From the theory of elliptic functions it is known that $\left(
k^{\prime}\right)  ^{2}$ should not touch the real axis which is cut from z=0
to z=-$\infty$ .This circumstance implies then that $k^{2}$ should not touch
the real axis cut from z=1 to z=+$\infty$ .Using Eq.(4.6) we obtain now
\begin{equation}
\frac{y_{1}}{y_{2}}=w(z)=\frac{K^{^{\prime}}(z)}{K(z)} \tag{4.17}%
\end{equation}
where $K^{^{\prime}}(z)$ is the same as $K$ in Eq.(4.16) with $k\rightarrow
k^{\prime}$ and $k^{\prime}$ being replaced by z $.$Strictly speaking,
solutions of Eq.(4.13) are determined with accuracy up to some constant
c.Therefore,the above result should be amended by
\begin{equation}
w(z)=c\frac{K^{^{\prime}}(z)}{K(z)}\equiv J^{-1}(z) \tag{4.18}%
\end{equation}
The choice of constant c is determined by the requirement that the above
mapping of z-plane (with cuts from 1 to $\infty$ and from -$\infty$ to 0)is
mapping onto the fundamental region of the Poincare H$^{2}$ plane as depicted
in Fig.s 2 and 3. If we choose c=i ,then the upper z half-plane is mapped into
the right half of the fundamental region D depicted in Fig.3 while the lower z
half plane is mapped into the left half of D [51]. The inverse of $w(z)$ is
given by the function $J(w)$ defined below:
\begin{equation}
z=J(w)=16q\prod\limits_{n=1}^{\infty}\left(  \frac{1+q^{2n}}{1+q^{2n-1}%
}\right)  ^{8} \tag{4.19}%
\end{equation}
with $q=exp$\{i$\pi w\}.$ These results were briefly mentioned in our earlier
work,Ref [14],e.g.see Appendix of Ref.[14]. Generalization of the obtained
results to the case of n punctures (n$\geq3)$is nontrivial (this is so because
the system of equations,Eq.(4.4) can be easily solved only for n=1 and
2,provided that the third point is located at $\infty).$Therefore$,$ many
methods can be used, in principle, with variable degree of success. In this
section we would like to discuss only methods which originate in string and
conformal field theories. In section 7 we shall discuss in some detail other possibilities.

In general, the accessory parameters $m_{k}=m_{k}(p_{1},...,p_{k})$ could be
some very complicated and,in general, unknown functions of puncture
locations.Some progress had been recently made based on the following key
observation made by Zograf and Takhtadjian[52]. \ Consider the Schwarzian
derivative, Eq.(4.3), rewritten in the following way
\begin{equation}
\{w,z\}=-\frac{1}{2}\left(  \partial_{z}\ln\left|  \frac{\partial w}{\partial
z}\right|  \right)  ^{2}+\frac{d^{2}}{dz^{2}}\ln\left|  \frac{\partial
w}{\partial z}\right|  \tag{4.20}%
\end{equation}
This result coincides with that given by Eq.(4.3) as can be seen by direct
calculation.At the same time,let $\varphi(z)=\ln\left|  \frac{\partial
w}{\partial z}\right|  $,then we can write instead of Eq.(4.20) the following
result
\begin{equation}
\{w,z\}=\frac{d^{2}}{dz^{2}}\varphi-\frac{1}{2}\left(  \frac{d\varphi}%
{dz}\right)  ^{2} \tag{4.21}%
\end{equation}
or,in view of Eq.(4.2),
\begin{equation}
\frac{d^{2}}{dz^{2}}\varphi-\frac{1}{2}\left(  \frac{d\varphi}{dz}\right)
^{2}=\frac{1}{2}\sum\limits_{k=1}^{n}\frac{1}{\left(  z-p_{k}\right)  ^{2}%
}+\sum\limits_{k=1}^{n}\frac{m_{k}}{z-p_{k}} \tag{4.22}%
\end{equation}
The function $\varphi$ satisfies the Liouville equation
\begin{equation}
\frac{\partial^{2}}{\partial z\partial\bar{z}}\varphi=\frac{1}{2}\exp
\varphi(z,\bar{z}) \tag{4.23}%
\end{equation}
which describes the surfaces of constant negative curvature $\kappa=-1.$The
Poincare H$^{2}$ model ,Eq.(3.10), equipped with such type of metric becomes a
model of the hyperbolic (or Lobachevski ) plane.The Liouville equation can be
obtained as equation of motion coming from the variation of the Liouville
action functional S[$\varphi]$ given by
\begin{equation}
S[\varphi]=\int\limits_{\Delta}d^{2}z\left[  (\frac{d\varphi}{dz}%
)^{2}+e^{\varphi}\right]  +counterterms \tag{4.25}%
\end{equation}
where $\Delta$ is determined by $\Delta=\Omega\backslash\cup_{i=1}%
^{n-1}\{\left|  z-p_{i}\right|  <r\}\cup\{\left|  z\right|  >\frac{1}%
{r}\},r\longrightarrow0^{+}$ and the explicit analytic form of the
counterterms is not important for what will follow next. By swithching from
the Lagrangian to the Hamiltonian formalism we obtain ,using Eq.(4.25) , the
energy-momentum (the stess-energy) tensor $T_{zz}$ given by
\begin{equation}
T_{zz}=\frac{d^{2}}{dz^{2}}\varphi-\frac{1}{2}\left(  \frac{d\varphi}%
{dz}\right)  ^{2}\equiv T \tag{4.26}%
\end{equation}
This result can now be combined with Eq.(4.22). \ This leads,in view of
eq.s(4.5), (4.21) and (4.22) to the result:
\begin{equation}
y^{^{\prime\prime}}+\frac{1}{2}Ty=0. \tag{4.27}%
\end{equation}
This equation represents the classical limit of the corresponding
quantum/statistical mechanical equation considered in the famous paper by
Belavin , Polyakov and Zamolodchikov (section 5 of Ref.[53]) in connection
with the development of the conformal field theory.Systematic corrections to
these classical results ,Eq.s(4.20)-(4.27), can be achieved through Feynman's
way of doing quantum mechanics.That is one formaly defines the path integral
\begin{equation}
<\Omega>=\int\limits_{C(\Omega)}D[\varphi]\exp\{-\frac{1}{2\pi h}S[\varphi]\}
\tag{4.28}%
\end{equation}
where $\Omega=S^{2}\backslash\{p_{1},...,p_{n}\}$ as before, C($\Omega)$ is
the appropriatly chosen domain of functional integration and $h$ is some
control parameter which is fixed at the end of calciulations[54].It is not our
purpose here to discuss the implications of such path integral formulation of
the uniformization problem.The details could be found in the references
already provided.Here we only are concerned with calculating the accessory
parameters $m_{k}.$These parameters can be obtained ,in principle,by means of
a simple looking formula[54,55]:
\begin{equation}
m_{k}=\frac{1}{2\pi}\frac{\partial S}{\partial p_{k}}. \tag{4.29}%
\end{equation}
The simplicity of this formula is somewhat mileading since it was obtained
with help of a saddle point approximation .And even within this approximation
actual computation can be made only for some special cases,e.g. when two
punctures are very close to each other while the rest are far away. In
particular, for n%
$<$%
4 ,one obtains [55]
\begin{equation}
m_{k}=-\frac{1}{2(p_{k}-p_{n})},\text{ \quad}p_{k}\text{ }\rightarrow p_{n},
\tag{4.30}%
\end{equation}
while for n$\geq4$ and one of the punctures is being located at zero ,one
obtains [],
\begin{equation}
m=-\frac{1}{2p}+\frac{\pi^{2}}{2p\left(  \ln\left|  p\right|  \right)  ^{2}%
}\text{ , \quad}p\longrightarrow0. \tag{4.31}%
\end{equation}
It is also possible to obtain the exact results for some very special
symmetric arrangement of the punctures.Whence,although, in principle, solution
of Eq.(4.5) solves the uniformizaton (or mapping) problem completely, in
practice,even if all $m_{k}$ would be known, solution of Eq.(4.5) could be so
complicated that its practical use in path integral calculations similar to
that described in section 2 becomes nonpractical.Nevertheless,our efforts are
not completely in vain.\ For example, we have learned already from example of
the trice punctured sphere that solution of Eq.(4.5) not only provides the
desired map but also gives the exact shape of the fundamental region in
H$^{2}$plane ,e.g.see Fig.s 2 and 3. In order to use other methods,e.g.see
section 7,we also need the precise shape of the fundamental region
or,equivalently,the explicit form of the unimodular transformation,Eq. (3.8).
For the fundamental domain depicted in Fig.3 the corresponding unimodular
transformation is known to be [51]
\[
a=\left\|
\begin{tabular}
[c]{cc}%
1 & 0\\
2 & 1
\end{tabular}
\right\|  \text{ \quad and }b=\left\|
\begin{tabular}
[c]{cc}%
1 & 2\\
0 & 1
\end{tabular}
\right\|  \text{ .}%
\]
in accord with earlier obtained Eq.(3.6) .Analogously, if the points are
arranged as n-th roots of unity , it can be shown [50], that
\begin{equation}
m_{k}=-\frac{1}{2}p^{-k} \tag{4.32}%
\end{equation}
and the corresponding unimodular transformation can be found explicitly as
well.Fortunately, this is not needed as results of section 7 indicate. The
most important for us is the fact that the results obtained in this section
for the trice punctured sphere could be used for unformization of the Riemann
surface of \textbf{any} genus .In this work we would like only to provide an
outline of these far reaching results, in section 7. In the meantime,in order
to keep things in proper perspective,we would like now to discuss some topics
which are directly related to just covered.

\section{$\quad\quad$Connections with the Riemann-Hilbert Problem and
Knizhnik-Zamolodchikov equations}

Eq.(4.5) with $\{w,z\}$ given by Eq.(4.2) is a typical case of Fuchsian
differential equation[].The most general form of equations of Fuchsian type is
known to be given by
\begin{equation}
y^{(p)}+q_{1}(x)y^{(p-1)}+\cdot\cdot\cdot+q_{p}(x)y=0 \tag{5.1}%
\end{equation}
where the coefficients $q_{i}(x)$ \textbf{near some singular point} $a$ are
given by
\begin{equation}
q_{i}(x)=\frac{r_{i}(x)}{\left(  x-a\right)  ^{i}}\text{ , \quad\quad
}i=1,...,p \tag{5.2}%
\end{equation}
where functions $r_{1}(x),...r_{p}(x)$ are holomorphic at $a.$ The above p-th
order linear differential equation has, in general, p independent solutions
.To find these solutions it is convenient to replace the above higher order
differential equation with the equivalent system of the first order
differential equations. This can be accomplished through the following changes
of variables :
\[
y=z^{1},
\]%

\[
(x-a)\frac{dy}{dx}=z^{2},
\]%

\[
\cdot\cdot\cdot\cdot\cdot\cdot\cdot\cdot\cdot\cdot\cdot\cdot\cdot\cdot
\]
\[
(x-a)^{p-1}\frac{d^{p-1}y}{dx^{p-1}x}=z^{p}.
\]
Using such defined z variables and the fact that Eq.(5.1) is equivalent to the
system of equations given below
\[
\frac{dy^{1}}{dx}=y^{2}
\]%

\begin{equation}
\frac{dy^{2}}{dx}=y^{3} \tag{5.3}%
\end{equation}%
\[
\cdot\cdot\cdot\cdot\cdot\cdot\cdot\cdot\cdot\cdot\cdot\cdot
\]
\[
\frac{dy^{p}}{dx}=-q_{p}(x)y^{1}-\cdot\cdot\cdot-q_{1}(x)y^{p}
\]
it is a simple matter to bring Eq.(5.1) to the canonical form
\begin{equation}
\frac{dz^{{}}}{dx}=A(x)z\text{ ,} \tag{5.4}%
\end{equation}
where the vector z is given by
\begin{equation}
z=\left[
\begin{array}
[c]{c}%
z^{1}\\
\cdot\\
\cdot\\
\cdot\\
z^{p}%
\end{array}
\right]  \tag{5.5}%
\end{equation}
and the matrix $A(x)$ is given by
\begin{equation}
A(x)=\frac{1}{x-a}\left\|
\begin{array}
[c]{cccccc}%
0 & 1 & 0 & 0 & \cdot\cdot\cdot & 0\\
0 & 1 & 1 & 0 & \cdot\cdot\cdot & 0\\
0 & 0 & 2 & 1 & \cdot\cdot\cdot & 0\\
\cdot\cdot\cdot & \cdot\cdot\cdot & \cdot\cdot\cdot & \cdot\cdot\cdot &
\cdot\cdot\cdot & \cdot\cdot\cdot\\
-r_{p} & -r_{p-1} & \cdot\cdot\cdot & \cdot\cdot\cdot & \cdot\cdot\cdot &
p-1-r_{1}%
\end{array}
\right\|  \tag{5.6}%
\end{equation}
It is clear that the above arguments can be extended to the case when ,instead
of one singularity located at $a$ we would have n singularities located at
$a_{i}$ , i=1-n.Then,instead of Eq.(5.4),we would have equation which looks
like Eq.(5.4) but with matrix $A(x)$ being replaced by
\begin{equation}
A(x)=\sum\limits_{i=1}^{n}\frac{1}{x-a_{i}}B_{i}(x) \tag{5.7}%
\end{equation}
with matrix $B_{i}(x)$ being nonsingular at above points.For example, the
hypergeometric equation, Eq.(4.13), can be presented in Fuchsian form given
below
\begin{equation}
\frac{dz}{dx}=\left(  \left(
\begin{array}
[c]{cc}%
0 & 0\\
-ab & -c
\end{array}
\right)  \frac{1}{x}+\left(
\begin{array}
[c]{cc}%
0 & 1\\
0 & c-(a+b)
\end{array}
\right)  \frac{1}{x-1}\right)  z \tag{5.8}%
\end{equation}
while the Bessel equation , Eq.(2.34), which upon trivial rescaling can be
brought to the standard form
\begin{equation}
\frac{d^{2}y}{dx^{2}}+\frac{1}{x}\frac{dy}{dx}+(1-\left(  \frac{\nu}%
{x}\right)  ^{2})y=0 \tag{5.9}%
\end{equation}
can be presented in the Fuchsian form of the following type
\begin{equation}
\frac{dz}{dx}=\frac{1}{x}\left(
\begin{array}
[c]{cc}%
0 & 1\\
\nu^{2}-x^{2} & 0
\end{array}
\right)  z \tag{5.10}%
\end{equation}
where the column vector z is given by
\begin{equation}
z=\left(
\begin{array}
[c]{c}%
y\\
x\frac{dy}{dx}%
\end{array}
\right)  . \tag{5.11}%
\end{equation}
Eq.(4.7) is also of Fuchsian type . It can be brought to from of Eq.(5.10)
with matrix $A(x)$ given by
\begin{equation}
A(x)=\frac{1}{x}\left(
\begin{array}
[c]{cc}%
0 & 1\\
-1 & 1
\end{array}
\right)  \tag{5.12}%
\end{equation}
with vector z being the same as in Eq.(5.11).

Consider now Eq.(5.4) from different perspective. First ,assume that there is
some matrix function $\Gamma(x)$ such that we can relate vector z to another
vector Z via
\begin{equation}
Z=\Gamma(x)z. \tag{5.13}%
\end{equation}
This then allows us to write
\[
\frac{dZ}{dx}=\frac{d\Gamma}{dx}z+\Gamma\frac{dz}{dx}%
\]
or,equivalently,
\begin{equation}
\frac{dz}{dx}=\Gamma^{-1}\frac{dZ}{dx}-\Gamma^{-1}\frac{d\Gamma}{dx}z.
\tag{5.14}%
\end{equation}
Substitution of this result back into Eq.(5.4) and elimination of Z with help
of Eq.(5.13) produces
\begin{equation}
\frac{dz}{dx}=A^{\prime}(x)z \tag{5.15}%
\end{equation}
with matrix $A^{\prime}(x)$ being given by
\begin{equation}
A^{\prime}(x)=\Gamma A\Gamma^{-1}+\frac{d\Gamma}{dx}\Gamma^{-1}. \tag{5.16}%
\end{equation}
At this point the reader can easily recognize the gauge transformation
characteristic of the nonabelian gauge field theories [56].Unlike the
field-theoretic case, our field is \textbf{nonrandom }and,therefore, we cannot
use field-theoretic methods which always involve the averages over randomly
fluctuating fields.Such averages make sense ,nevertheless,since in physical
reality the punctures(poles) are not fixed but can move(see,however,section 7
for further details).. One may think of Eq.(5.4) (or (5.15)) as some sort of
equation for the massless Dirac particle(e.g.neutrino) placed in
time-dependent (or space-dependent but not both!)nonabelian gauge field
[56].The question arises: can one consider instead the \textbf{massive} Dirac
particle in such gauge field ? The answer turns out to be negative as we are
going to demonstrate momentarily this purpose instead of Eq.(5.7) we would
like to consider more general matrix given by \quad%
\begin{equation}
A(x)=\sum\limits_{i=1}^{n}\frac{1}{x-a_{i}}B_{i}(x)+B(x) \tag{5.17}%
\end{equation}
where the matrix $B(x)$ is not singular for all x . Let all $a_{i}$
$\neq\infty$ and require our system of equations ,Eq.(5.7) to be nonsingular
at $\infty$ as well. This puts some constraint on both $B_{i}(x)$ and
$B(x).$By replacing $x$ with $\tau=x^{-1}$in Eq.s (5.7),(5.17) a simple
analysis shows that $B(x)=0$ and,in addition,
\begin{equation}
\sum\limits_{i=1}^{n}B_{i}(x)=0 \tag{5.18}%
\end{equation}
Accordingly,we cannot use massive Dirac particle interpretation of the
obtained results. At the same time, let $x\in\Omega$ where\textbf{\ }$\Omega$
was defined at the begining of section 4.Let ,as before,$H^{2}$ be the
universal covering space of $\Omega$ (that is we are considering at least
three punctures).By analogy with section 3 ,let $G$ be the group of motions in
the covering space. Let now $\sigma$ and $\delta\in G$ ,then if $z(x)$ is
solution of the system Eq.(5.7) and $z(\tilde{x})$ is solution in the covering
space,then as before, $z(\sigma\tilde{x}),$is also a solution to Eq.(5.7) in
the covering space .Moreover, in view of the linearity of Eq.(5.4) ,such
solution is defined only with accuracy up to some \textbf{constant} matrix
$\chi(\sigma).$ Let us introduce notations:$z(\sigma\tilde{x})=z\circ\sigma$
and $z(\sigma\tilde{x})\chi(\sigma)=\left(  z\circ\sigma\right)  \chi
(\sigma).$ Using these results,we obtain as well
\[
z\circ\delta=[\left(  z\circ\sigma\right)  \chi(\sigma)]\circ\delta=\left(
z\circ\sigma\circ\delta\right)  \chi(\sigma)
\]
This naturally implies
\[
\left(  z\circ\delta\right)  \chi(\delta)=\left(  z\circ\sigma\circ
\delta\right)  \chi(\sigma)\chi(\delta)=\left(  z\circ\left(  \sigma
\delta\right)  \right)  \chi(\sigma\delta)
\]
that is
\begin{equation}
\chi(\sigma\delta)=\chi(\sigma)\chi(\delta) \tag{5.19}%
\end{equation}
which is a \textbf{monodromy} group condition analogous to that we have
already encountered in connection with Eq.(2.7). In case of the trice
punctured sphere, using Eq.(3.1), we obtain
\begin{equation}
\chi(abc)=\chi(a)\chi(b)\chi(c)=\chi(1)=1 \tag{5.20}%
\end{equation}
Evidently,in case of more than 3 punctures we should have
\begin{equation}
\prod\limits_{i=1}^{n}\chi(a_{i})=1 \tag{5.21}%
\end{equation}
where $a_{i}$ is the generator of motion around i-th loop (e.g.see Fig.1).The
classical Riemann-Hilbert (R-H) problem can be formulated now as follows.

\textbf{R-H} \textbf{problem}:\textit{\ given location of the points }%
$p_{1},...,p_{n}\in S^{2}$\textit{\ and matrices }$\chi(a_{i}),$\textit{\ find
a linear differential equation of the type given by Eqs(5.4)-.(5.7) whose
monodromy group coincides with the group associated with matrices }$\chi(a_{i}).$

If we allow points $p_{1},...,p_{n}$ to move, then one may pose

\textbf{The related problem}: \textit{find conditions under which for the
fixed monodromy group (and this is very natural since topologically nothing
changes) solutions of Eq.(5.4) and the matrix }$A(x)$\textit{\ will depend
continuously on the set }$p_{1},...,p_{n}.$\textit{\ }

More precisely, let $z=z(x_{0};x)$ be solution of the system of Eqs(5.4)
,where $x_{0}=\{p_{10},...,p_{n0}\}$ .Then, the necessary and sufficient
condition for the matrix $A(x)$ as function of parameters $p_{1},...,p_{n}$
and $x_{0}$ to be associated with the \textbf{fixed} monodromy group depends
upon the solvability of the following set of (consistency) equations
\[
\frac{\partial A_{j}}{\partial p_{i}}=[A_{j},A_{i}](\frac{1}{p_{i}-p_{j}%
}-\frac{1}{p_{i0}-p_{i}})\text{ , }j\neq i,
\]%
\begin{equation}
\frac{\partial A_{i}}{\partial p_{i}}=-\sum\limits_{j\neq i}[A_{i},A_{j}%
]\frac{1}{p_{i}-p_{j}}\text{ ,} \tag{5.22}%
\end{equation}%
\[
\frac{\partial A_{i}}{\partial x_{0}}=\sum\limits_{j\neq i}[A_{i},A_{j}%
]\frac{1}{x_{0-}p_{j}}\text{ , }i=1,...,n\text{ .}%
\]
The system of equations given above is known in the literature as
Schlesinger's equations [57].Study of solutions of these equations is directly
associated with study of the exactly integrable classical and quantum systems
[58] , conformal field theory[59] and Einsteinian 3+1 gravity[60] and
,therefore, is not going to be discussed further in this section.

Nevertheless,to make our presentation self-contained, we would like to discuss
briefly connections between the system of Eqs.(5.7) and that known in the
literature as Knizhnik-Zamolodchikov (K-Z) equations [61].The system of
equations (5.4) can be formally solved using method of Lappo-Danilevsky [62].
Let the matrix of solutions of Eq.(5.4) be just a unit matrix I for $x=x_{0}%
,$then , for $x\neq x_{0}$ we can formally write solution to Eq.(5.4) in the
following form
\begin{equation}
z(x)=I+\sum\limits_{\nu=1}^{\infty}\sum\limits_{j_{1},...,j_{\nu}}%
^{1,...,n}L_{x_{0}}(p_{j_{1}},...,p_{j_{\nu}}\mid x)A_{j_{1}}\cdot\cdot\cdot
A_{j_{\nu}} \tag{5.23}%
\end{equation}
where the functions $L_{x_{0}}(...)$ are defined recursively as follows
\begin{equation}
L_{x_{0}}(p_{j_{1}}\mid x)=\int\limits_{x_{0}}^{x}\frac{dy}{y-p_{j_{1}}}
\tag{5.24}%
\end{equation}
and,accordingly,
\begin{equation}
L_{x_{0}}(p_{j_{1}},...,p_{j_{\nu}}\mid x)=\int\limits_{x_{0}}^{x}%
\frac{L_{x_{0}}(p_{j_{1}},...,p_{j_{\nu}-1}\mid y)}{y-p_{j_{\nu}}}dy.
\tag{5.25}%
\end{equation}
Using these results, let us consider now the following system of equations
\begin{equation}
\frac{\partial f}{\partial x_{i}}=h\sum\limits_{j=1,j\neq i}^{n}\frac{t_{ij}%
}{x_{i}-x_{j}}f(x_{1},...,x_{n}). \tag{5.26}%
\end{equation}
Here $h$ is some prescribed constant(control parameter) and the symmetric
matrix $t_{ij}$ is also assumed to be given (actually,the matrix $t_{ij}$ is
\textbf{not the matrix }but rather the tensor product of two matrices[61] but
for the moment we shall ignore this fact ).The above system of equations is
known in the literature as K-Z system of equations [61].It has numerous
physical applications ,e.g.see Ref.[14] ,which are not going to be discussed
here.What will be important for us is its connection with Eqs(5.4) and
(5.22).To establish this connection several steps are needed. First,using the
property of symmetry of the matrix $t_{ij}$ it is sraightforward to obtain the
following auxiliary equations
\begin{equation}
\sum\limits_{i=1}^{n}\frac{\partial f}{\partial x_{i}}=0\text{ and }%
\sum\limits_{i=1}^{n}x_{i}\frac{\partial f}{\partial x_{i}}=\sum\limits_{1\leq
i<j\leq n}^{{}}f(x_{1},...,x_{n}) \tag{5.27}%
\end{equation}
Eq.s(5.27) reduce the number of independent variables in K-Z equation to $n-2$
.In particular, solution $f(x_{1},x_{2},x_{3})$ of K-Z Eq.(5.26) actually
depends only upon one variable which we shall denote as $x.$More specifically,
if we look for solution of Eq.(5.26) in the form
\begin{equation}
f(x_{1},x_{2},x_{3})=(x_{3}-x_{1})^{h(t_{12}+t_{23}+t_{31})}g(x) \tag{5.28}%
\end{equation}
where
\[
x=\frac{x_{2}-x_{1}}{x_{3}-x_{1}}\text{ ,}%
\]
then,substitution of such an ansatz into Eq.(5.26) produces the following
equation for the function $g(x)$:
\begin{equation}
\frac{dg}{dx}=h(\frac{t_{12}}{x}+\frac{t_{23}}{x-1})g(x). \tag{5.29}%
\end{equation}
In the original paper by Knizhnik and Zamolodchikov[61] it is explained why
$t_{12}$and $t_{23}$ are actually 2$\times2$ matrices (not just usual numbers
which are components of the matrix).Comparison with Eq.(5.8) then suggests
immediately that we are dealing with the hypergeometric equation and ,indeed,
Knizhnik and Zamolodchikov had obtained two solutions for $g(x)$ in terms of
hypergeometric functions .Their solution is not the most general ,however. A
complete solution was found by Drinfeld[63] and ,more explicitly,by Le and
Murakami [64].It is noteworthy to describe it since it is related to quantum
groups and Vassiliev invariants for knots and links [65-67].To describe this
solution, let us consider instead of Eq.(5.29) the following generalization
\begin{equation}
\frac{dg}{dx}=\frac{h}{2\pi\sqrt{-1}}(\frac{A}{x}+\frac{B}{x-1})g(x)
\tag{5.30}%
\end{equation}
where the prefactor $\frac{h}{2\pi\sqrt{-1}}$ is chosen for convenience only
(since $\frac{h}{2\pi}$ is just the usual Planck constant in quantum
mechanics)and the factors $A$ and $B$ are some non-commuting variables,
e.g.matrices.Let $g_{1}(x)$ and g$_{2}(x)$ be two independent solutions of
Eq.(5.30).Then, as it was shown by Drinfeld [63], g$_{1}(x)=g_{2}(x)\Phi(A,B)$
where the function $\Phi$ of two noncommuting variables is known in the
literature as Drinfeld associator .The explicit form of $\Phi$ is hard to find
in closed form .If h is small parameter,then ,after some tedious
calculations[64]which involve use of Lappo-Danilevskii-type expansions ,one
finally obtains
\begin{equation}
\Phi(A,B)=1-\frac{\varsigma(2)}{\left(  2\pi\sqrt{-1}\right)  ^{2}}%
[A,B]h^{2}+\frac{\varsigma(3)}{\left(  2\pi\sqrt{-1}\right)  ^{3}%
}([[A,B],B]-[A,[A,B]])h^{3}+... \tag{5.31}%
\end{equation}
Here $\varsigma(x)$ is Riemann zeta function.To obtain solutions $g_{1}(x)$
and $g_{2}(x)$ is rather easy in the neigbourhoods of 0 and 1.To this purpose
it is sufficient to look for solution of $g_{1}$in the form
\begin{equation}
g_{1}(x)=P(x)x^{hA} \tag{5.32}%
\end{equation}
and,analogously,
\begin{equation}
g_{2}(x)=Q(1-x)(1-x)^{hB} \tag{5.33}%
\end{equation}
Assuming that functions $P(x)$ and Q(1-x) can be represented in terms of power
series expansions it is rather straigthforward to find the successive
coefficients.For values of x not necessarily close to x=0 and x=1 again one
should be looking for solution using Lappo-Danilevsky method. Altschuler and
Freidel had demostrated [66] how Drinfeld associator $\Phi$ can be used for
calculation of the regularized Vassiliev-Kontsevich invariant for knots and
links.It is not our purpose to reproduce these calculations in this
paper.Instead,we would like to mention yet another physical application of the
obtained results.

To this purpose, let us consider again system of equations Eq.(5.4) with
matrix $A(x)$ being given by Eq.(5.7).Let us first choose matrices B$_{i}$ to
be constants. Then ,the gauge transformation ,Eq.(5.16), is just a similarity
transformation so that we can always select such $\Gamma^{\prime}s$ that the
resulting matrix be diagonal. Without loss of generality ,we shall assume that
this is the case from now on. Let us then look for a solution of Eq.(5.4) in
the vicinity of one of the points, say, $a_{j}.$ We can always choose the
system of coordinates located at this point.Then, in the vicinity of this
point we essentially have to solve the following equation
\begin{equation}
\frac{dz}{dx}=\frac{A_{j}}{x}z \tag{5.34}%
\end{equation}
The solution of this equation is obtained at once and is given by $z=x^{A_{j}%
}$ =$\exp\{A_{j}\ln x\}.$ Now, if x is in the complex plane (or on S$^{2})$
use of Eq.(2.16) produces ,after one rotation around x=0, the following
solution $z\circ\sigma=\exp\{2\pi iA_{j}\}z\equiv\chi(\sigma)z.$ If now the
matrix $A(x)$ is x-dependent,then one can look for solution of Eq.(5.4) in the
form
\begin{equation}
z(x)=\sum\limits_{j=1}^{n}\Phi_{j}(x)(x-a_{j})^{A_{j}} \tag{5.35}%
\end{equation}
where $A_{j}$ is the value of A(x) at x=$a_{j}$ (surely ,upon proper
diagonalization) and the functions $\Phi_{j}(x)$ are holomorphic at x=$a_{j}$
.Instead of S$^{2},$let us consider a disc D$^{2}$ first. Let x$_{0}$ be the
location of singular point(s) in D$^{2}$ and let $\gamma_{1},...,\gamma_{n}$
be a set of loops which go around the singular points while \{ $\chi
(\sigma_{j})\}$ be the respective monodromy matrix set.Now we cover D$^{2}$ by
domains $U_{j}$ so that each domain $U_{j}$ has only one singularity $a_{j}$
.Using Eq.(5.35) in each domain we have z$_{j}=\Phi_{j}(t)t^{A_{j}}$ (where t
is the local coordinate in U$_{j}^{\prime}$th domain) so that at the
intersection of $U_{i}$ and $U_{j}$ domains we should have the consistency
condition
\begin{equation}
\Phi_{j}^{-1}\Phi_{j+1}=t^{A_{j}}t^{-A_{j+1}}\mid U_{j}\cap U_{j+1} \tag{5.36}%
\end{equation}
or ,more generally, if one considers S$^{2},$%
\begin{equation}
\Phi_{j+}(x)=G(x)\Phi_{j-}(x) \tag{5.37}%
\end{equation}
with G(x) being some known function . This type of boundary value problem is
very well known in the theory of elasticity [68] and was thoroughly studied
since it was first formulated by Hilbert in 1900 (21st Hilbert's problem)
[69].Whence, the use of the non -Abelian gauge field theories has a long
history . Birkhoff [70] had shown that solution of the Hilbert problem
,Eq(5.37) is equivalent to the solution of the Fuchsian system of equations ,
Eq.(5.4). More readable account of this proof is contained in Ref.[71] .Use of
Riemann-Hilbert problem in the the theory of exactly integrable equations is
well documented,e.g. in Ref[72].Accordingly, it is not our intention to
discuss these topics in this paper.Instead we would like to discuss topics
related to random walks on groups in the following section.

\section{Random walks on free groups: an introduction}

By now,it should become clear that even for 3 punctures in the plane the path
integral approach developed in section 2 becomes nonpractical.In current
literature the situation with 3 or more punctures is treated with help of an
auxiliary abelian or nonabelian gauge fields as described in our Phys.Reports
,Ref[14]. Such approach though plausible but mathematically less rigorous. To
realize the difficulty , we would like to mention the following. In Ref.[7]
the trice punctured sphere (or twice punctured plane) problem was solved with
help of path integral methods.However,the major question about the
recurrence/transience was not even addressed. Moreover,the method used in
Ref.[7] does not admit generalization to the multiple punctured case .
Accordingly, it is necessary to develop other methods of dealing with these
problems. In this section we shall discuss an alternative approach which
originates from the fundamental work of Kesten[73] .

In section 3 we had discussed presentation for the group $G$,e.g.see Eq.(3.2)
(with all $R_{i}=0)$ and associated with it word W ,Eq.(3.4).Following Kesten,
let us consider the random walk on $G$ in which every step consist of right
multiplication by a$_{i}$ (or its inverse) each with probability p$_{i}%
.$Surely,one can complicate matters by assigning different probabilities for
direct and inverse elements but we shall not complicate our presentation at
this time. For the\textbf{\ free} group of n elements we should require at
each step
\begin{equation}
2\sum\limits_{i=1}^{n}p_{i}=1 \tag{6.1}%
\end{equation}
The random walk thus defined determines a Markov chain whose possible states
are elements of G. The transition probability(to be precisely defined below)
from some $W_{1}\in G$ to another $W_{2}\in G$ is given by the probability
that $W_{2}$ is reached in one step from W$_{1}.$ So far all this looks very
detached from the previous discussions. To make a connection,we shall
,following Ref.[3] , introduce some additional concepts.Let X be a metric
space in which the distance (the geodesic distance) d(x,y) between points x
and y which belong to X is formally determined by
\begin{equation}
d(x,y)=\left|  x-y\right|  \tag{6.2}%
\end{equation}
In particular, if $W_{1},W_{2}\in G$ ,then
\begin{equation}
\left|  W_{1}-W_{2}\right|  =\min\{m\text{ }\left|  W_{1}^{-1}W_{2}=a_{1}%
a_{2}...a_{m}\right.  \}. \tag{6.3}%
\end{equation}
That is the distance between words is determined my the minimal word.For
example, let $W_{1}=a_{1}a_{2}a_{2}^{-1}a_{3}$ .Surely ,this word is not
minimal since the combination $a_{2}a_{2}^{-1}$=1 can be easily removed. As
simple as it is, the general problem of comparing different words ,known as
word problem [44] , cannot be solved in general[74]. This does not mean that
the problem cannot be solved for a specific group,for example, for the free
group.More on this subject is discussed in the classical
monograph,Ref.[44].More important for us is the fact that such defined
distance,Eq.(4.3), possesses the property of translational invariance,just
like in the case of Eq.(2.3).Indeed, let us consider $\left|  \gamma
W_{1}-\gamma W_{2}\right|  .$According to Eq.(6.3) we have
\begin{equation}
\left|  \gamma W_{1}-\gamma W_{2}\right|  =\left(  \gamma W_{1}\right)
^{-1}\gamma W_{2}=W_{1}^{-1}\gamma^{-1}\gamma W_{2}=W_{1}^{-1}W_{2}. \tag{6.4}%
\end{equation}
It is possible to introduce not only the distance but even the curvature of
the group,etc.For instance,following Milnor[75], for a finitely generated
group, i.e.group of finite number of generators,e.g. n , one can define the
growth function $\gamma(s)$ which for each positive integer s determines the
number of words with length $\leq s.$ Then, it can be shown , that for the
free group of n generators the growth function is given by
\begin{equation}
\gamma(s)=\left(  n(2n-1)^{s}-1\right)  /(n-1) \tag{6.5}%
\end{equation}
and this result is directly related to the curvature of the underlying
manifold whose fundamental group consist of n elements.

It should be noted,that the distance defined above is not the only possibility
in groups. For instance,if we fix some element,say unity, as a reference
point, then ,one can introduce another distance between x and y ,also known in
the literature as Gromov product[3,76,77], e.g.
\begin{equation}
d_{G}(x,y)=\frac{1}{2}(\left|  x\right|  +\left|  y\right|  -\left|
xy^{-1}\right|  ) \tag{6.6}%
\end{equation}
so that,by construction,$d_{G}(x,x)$=$\left|  x\right|  \geq0$ which makes
perfect sence.Using this result one can go further by providing the following

\textbf{Definition 6.1}.\textit{Let }$\delta$ \textit{be nonnegative real
number.The metric space X is said to be} $\delta-$\textit{hyperbolic if}
\begin{equation}
d_{G}(x,y)\geq\min(d_{G}(x,z),d_{G}(y,z))-\delta\tag{6.7}%
\end{equation}
\textit{for every x,y and z}$\in X.$\textit{Moreover, the metric space is
\textbf{hyperbolic}(in the sence discussed in sections 3 and 4) if there is a
real number }$\delta$\textit{\ so that X is }$\delta-$\textit{hyperbolic.}

It can be demonstrated[3,76,77] that:

a) every graph is $0-$ hyperbolic;

b)every free group of finite rank is hyperbolic.

Moreover, following Ref.[77], the $\delta-$\textbf{ultrametric} space X can
now be defined via$\ $%
\begin{equation}
\left|  x-y\right|  \leq\max(\left|  x-z\right|  ,\left|  y-z\right|
)+\delta\tag{6.8}%
\end{equation}

for all x,y,z $\in X$ .Using such definition, the following theorem can be proven[77]

\textbf{Theorem} 6.2. \textit{Every metric space X which satisfy the above
inequality is actually }$\delta-$\textit{hyperbolic.}

This means that the \textbf{ultrametric space is a special case of hyperbolic
space}. This fact could be used for solving some problems related to spin
glasses and other systems which require use of replicas[4].

Going back to our problem of symmetric random walk on free groups we need to
introduce two probabilities :

$m^{(n)}=$ Probability of returning to the unit element $e$ at n-th step given
that one initially starts at $e.$

$r^{(n)}=$Probability of returning for the \textbf{first time} to $e$ at n-th
step at n-th step given that one initially starts at $e$.

With such defined probabilities the generating functions $m(x)$and $r(x)$ can
be now formally introduced as follows:
\begin{equation}
m(x)=\sum\limits_{n=0}^{\infty}m^{(n)}x^{n}\text{ , provided that }m^{(0)}=1
\tag{6.9}%
\end{equation}
and
\begin{equation}
r(x)=\sum\limits_{n=1}^{\infty}r^{(n)}x^{n}. \tag{6.10}%
\end{equation}
As it was demonstrated by Feller[78],e.g.see page 311 of Feller's book.
\begin{equation}
m(x)=\frac{1}{1-r(x)}. \tag{6.11}%
\end{equation}
Thus,if $r(x)$ is known, $m(x)$ is known as well.To determine $r(x)$ one needs
to determine $r^{(n)}$ explicitly. Elegant solution of this combinatorial
problem is presented in Kesten's paper[73],e.g see pages 348,349 of Ref.[73]
and take into account similar derivations in Feller's book, Ref[78], with the
following result:
\begin{equation}
r(x)=\frac{n-(n^{2}-(2n-1)x^{2})^{\frac{1}{2}}}{2n-1}. \tag{6.12}%
\end{equation}
Use of Eq.(6.11) provides us with the following value of $m(x)$%
\begin{equation}
m(x)=\frac{2n-1}{n-1+(n^{2}-(2n-1)x^{2})^{\frac{1}{2}}} \tag{6.13}%
\end{equation}
with n being the number of generators in the free group $G$.

Now, we would like to use all this information in order to discuss the random
walk on Bethe lattice ,Fig.6,%
\begin{figure}
[ptb]
\begin{center}
\includegraphics[
height=2.2874in,
width=1.8317in
]%
{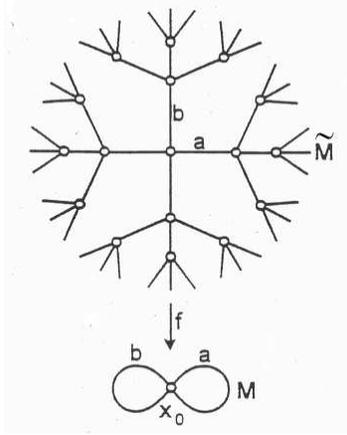}%
\caption{Figure 8 and its universal covering space}%
\end{center}
\end{figure}
which is just the universal covering space of figure eight (in the case of
free nonabelian group of two generators) or, more generally, the covering
space of the union of n circles[74] (in case of n generators).Such problem was
solved by conventional methods in Ref.[79] and we would like to borrow a
couple results from this reference for the sake of comparison.In particular,
let $P_{t}(l$ $\left|  m)\right.  $be the probability that the walk originated
at position $m$ will end up at position $l$ after $t$ steps.In case of the
Bethe lattice the position $m$ (or $l$) is determined with respect to the
origin, e.g.see Fig.6 ,so that $m$ means the number of links making a path
between the origin and point $m$,etc .The Markov chain equation is given by
\begin{equation}
P_{t+1}(l\left|  m\right.  )=\sum\limits_{l^{^{\prime}}}\gamma(l,l^{\prime
})P_{t}(l^{\prime}\left|  m)\right.  \tag{6.14}%
\end{equation}
with $\gamma(l,l^{\prime})$ being determined by
\begin{equation}
\gamma(l,l^{\prime})=\left\{
\begin{array}
[c]{c}%
(1-\frac{1}{2n})\delta_{l,l^{\prime}}+\frac{1}{2n}\delta_{l,l^{\prime}%
-1}\text{ ,}l^{\prime}\geq1\\
\delta_{l,l^{\prime}+1}\text{ \ \ \ \ \ \ \ \ \ \ \ \ \ \ \ \ , }l^{\prime
}=0\ \text{\ \ \ \ \ \ \ }%
\end{array}
\right.  \tag{6.15}%
\end{equation}
The probability $\Pi(0)$ of returning to the origin ,defined in Eq.(2.45), is
calculated to be
\begin{equation}
\Pi(0)=\frac{1}{2n-1} \tag{6.16}%
\end{equation}
and is surely less than one for n$\geq$2. Hence, for n$\geq2$ the walk is
transient and this result is in accord with results obtained by Lyons and
McKean[19] and McKean and Sullivan[20] who used different methods . For n=1
the walk is recurrent, since $\Pi(0)=1,$ and this result is in accord with
earlier obtained in section 2 for 2 punctures one of which being located at
infinity.We would like now to reproduce this result using the theory of random
walks on groups.This is needed for various reasons discussed in the
Introduction ,below, and to be discussed in the next section.To begin,
following Kasteleyn[23] ,we consider an analogue of Eq.(6.14) for group
elements(letters) $a_{1},...,a_{n}$ of some free group $G$ . Let $P_{t}(W)$ be
the probability of creating the word $W$ of t letters starting from $e.$ The
Markov chain equation analogous to Eq.(6.14) can now be written as
\begin{equation}
P_{t}(W^{^{\prime}})=\sum\limits_{W\in G}P_{t-1}(W^{-1}W^{^{\prime}%
})p(W)\text{ ,}t\geq1. \tag{6.17}%
\end{equation}
with $p(W)$ being defined as \textbf{stepping} probability which is the
probability of choosing a particular branch at each vertex of the tree.
Looking at Fig.6 ,we can easily notice that it is exactly the same probability
as for crossing the vertex of figure 8 (in case of two generator group).
Surely,one can complicate matters by assigning different weights to different
choices of crossing the vertex as it is done ,for example, in the case of the
Ising model. Such possibility immediately connects the walks on groups with
various statistical mechanical models as discussed ,for example, in Ref.[80].
We shall in this work restrict our discussion to the simplest possible case
,e.g. $p(W)=1/2n$ (motivated by Fig.6).As in the case of usual random
walks[42], it is convenient to introduce the generating function $P(W,z)$ as
follows
\begin{equation}
P(W,z)=\sum\limits_{t=0}^{\infty}P_{t}(W)z^{t} \tag{6.18}%
\end{equation}
This generating function obeys an equation of the following type
\begin{equation}
P(W^{^{\prime}},z)-\delta_{W^{^{\prime}},e}=z\sum\limits_{W\in G}%
P_{t-1}(W^{-1}W^{^{\prime}},z)p(W) \tag{6.19}%
\end{equation}
so that the probability $\Pi(W)$ that a walk starting at $e$ will ever visit
element $W$ (excluding the start as visit to $e$) is given by
\begin{equation}
\Pi(W)=\lim_{z\rightarrow1}\frac{P(W,z)-\delta_{W,e}}{P(e,z)} \tag{6.20}%
\end{equation}
The crucial moment now is to notice that ,actually, $P(e,z)=m(z)$ where $m(z)$
is given by Eq.(6.9).At this moment Kasteleyn claims that this observation
allows one to determine $P(W,z)$ with help of Eq.(6.19). Since the details of
the derivation are not provided in Kasteleyn's paper, we restore these
calculations in this work. To this purpose, let us assume that solution could
be sought in the form
\begin{equation}
P(W,z)=\lambda^{\left|  W\right|  }P(e,z) \tag{6.21}%
\end{equation}
with $\left|  W\right|  $ being defined by Eq.(6.3).\ Using these results in
Eq.(6.19) we obtain,
\begin{equation}
\lambda^{\left|  W^{^{\prime}}\right|  }P(e,z)-\delta_{W^{^{\prime}},e}%
=\frac{z}{2n}\sum\limits_{W\in G}\lambda^{\left|  W^{-1}W^{^{\prime}}\right|
}P(e,z). \tag{6.22}%
\end{equation}
Let us notice now that the actual length of the combination $\left|
W^{-1}W^{\prime}\right|  $ is equal to one since in one step of the random
walk only one letter is added or subtracted according to the definitions made
above.Finally, let $W^{^{\prime}}=e$ and ,since $\left|  e\right|  =0,$ we
obtain from Eq.(6.22) the following result:
\begin{equation}
\frac{P-1}{Pz}=\lambda\tag{6.23}%
\end{equation}
since
\begin{equation}
\sum\limits_{W\in G}=2n \tag{6.24}%
\end{equation}
by construction.Here $P=P(e,z).$By combining Eq.(6.13) with Eq.(6.23) we
obtain
\begin{equation}
\lambda=\frac{z}{n+(n^{2}-(2n-1)z^{2})^{\frac{1}{2}}} \tag{6.25}%
\end{equation}
Obtained result coincides \ exactly with that obtained by
Kasteleyn,Ref[23],e.g.see his Eq.8 , where it was given without derivation.By
combining Eqs(6.21) and (6.25) we obtain the final result for $P(W,z)$ which
can be used now for calculation of $\Pi(W).$In particular,for $W=e$ we
obtain,using Eq.(6.20),
\begin{equation}
\Pi(e)\equiv\Pi(0)=\frac{1}{2n-1} \tag{6.26}%
\end{equation}
in complete accord with earlier obtained Eq.(6.16) . It should become clear by
now,that consideration of random walks on groups provides practically as much
information as one could possibly get by much more tedious calculations using
the real-space approach. This is so because the group-theoretic \ methods
provide an equivalent and complete description of the underlying geometric and
topological properties of a given manifold[3,75-77].We would like now to
elaborate this observation in the next section.

\section{Some applications}

The recurrence/transience of random walks has actually numerous physical
applications which had been mentioned already in our previous works[14]. To
make our presentation self-contained,we would like to remind to our readers
about some of these applications and to discuss additional physical
applications in this section.

\subsection{Polymers}

First, the results of previous section can be used in the theory of polymer
solutions[14] since they provide the exact mathematical formulation of the
concept of entanglement through use of mathematically well defined concepts of
recurrence and transience. It should be clear by now that the random walks on
groups are\textbf{\ not at all} limited to the planar problems since the
homotopy of paths are not limited to two dimensions. In particular, one can
think now about random walks in the presence of knots and links(the
possibility mentioned already in our earlier work, Ref[14]). From this
reference we know that almost all flexible polymers are either knotted or
quasiknotted in solution in case if their length L is large. Formation of
knots/links may affect both static and dynamic properties of polymer solutions
and also polymer interfaces as is well known. Nevertheless, to detect the
presence of knots in solution is very nontrivial problem. Fortunately, using
methods developed in this work this problem may be solved to some extent. Let
us explain briefly how this can be done on examples of trefoil ,Fig.7,%
\begin{figure}
[ptb]
\begin{center}
\includegraphics[
height=2.2857in,
width=2.3315in
]%
{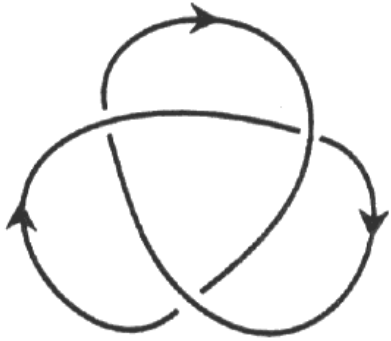}%
\caption{Planar projection of the trefoil knot}%
\end{center}
\end{figure}
and figure eight ,Fig.8,%
\begin{figure}
[ptbptb]
\begin{center}
\includegraphics[
height=2.2866in,
width=2.2866in
]%
{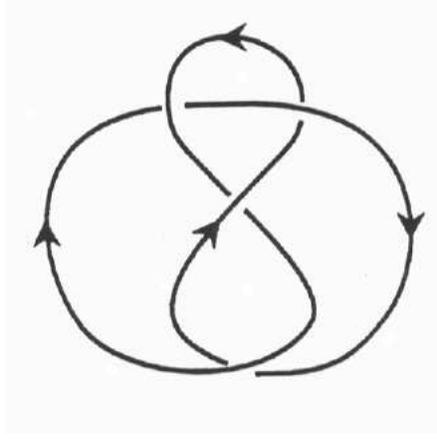}%
\caption{Planar projection of the figure 8 knot.}%
\end{center}
\end{figure}
knots.Using either results of our Appendix A2 of Ref[14] or more specialized
books on knot theory,e.g. Ref[43],[81], one obtains the following group
presentations
\begin{equation}
G_{T}=<a,b,c\left|  cb=ba=ac>\right.  \tag{7.1}%
\end{equation}
for the trefoil and
\begin{equation}
G_{8}=<a,b\left|  {}\right.  > \tag{7.2}%
\end{equation}
for the figure 8 knot complement groups.We would like to remind our readers
that the above group presentations are for the fundamental groups of the
\textbf{complements} of these knots in S$^{3}$(i.e.for S$^{3}\backslash K$
where $K$ is the knot of given type and S$^{3}=R^{3}\cup\{\infty\}).$It is
being assumed that the knot has finite thickness (framing) so that the
toroidal-like tunnel occupied by $K$ is removed from $R^{3}$ thus converting
it to space with possibly very complicated topology .The Cayley graph for the
fundamental group of the trefoil is depicted in Fig.9%
\begin{figure}
[ptbptbptb]
\begin{center}
\includegraphics[
height=2.2857in,
width=3.4316in
]%
{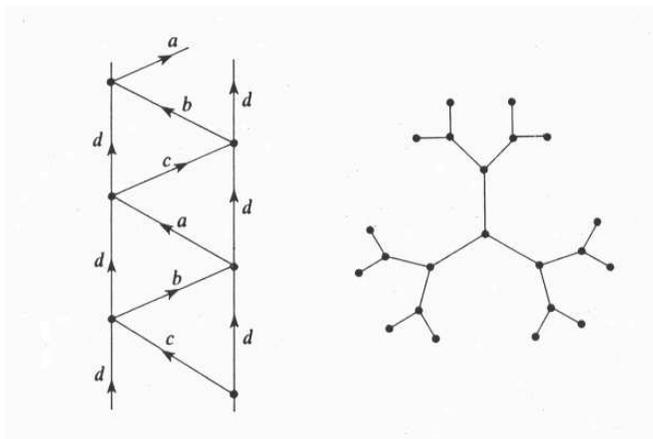}%
\caption{Cayley graph for the complement of the trefoil.Here d=cb.}%
\end{center}
\end{figure}
while that for the figure 8 is depicted in Fig.6. Fig.9 has rather complicated
structure which was discovered by Max Dehn [82] whose Dehn's twists we also
going to discuss below.When viewed from above ,one can see the three-valent
Bethe lattice(the terminology used in physics literature for the Cayley
graph[74]) . At each vertex of such lattice three infinite ladders are glued
together in an obvious way. To describe random \ walks on such lattices is not
an easy task and, besides,the presentation given by Eq.(7.1) is \textbf{not}
unique, for example,using results of Ref[81] one obtains as well.
\begin{equation}
G_{T}=<x,y\left|  x^{2}=y^{3}=1>\right.  \tag{7.3}%
\end{equation}
with the corresponding Cayley graph (part of it) depicted in Fig.10.%
\begin{figure}
[ptbptbptbptb]
\begin{center}
\includegraphics[
height=2.2874in,
width=4.5558in
]%
{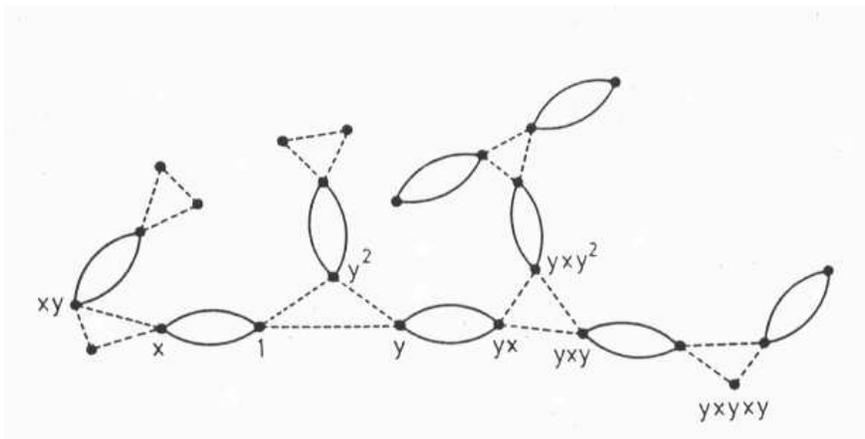}%
\caption{An alternative universal covering space for the complement of trefoil
knot.}%
\end{center}
\end{figure}
Clearly, the random walk in the complement of the figure 8 knot is transient
(that is if one could make a solution of linear and circular knotted polymers
which all are figure 8 knots ,the linear polymers will become entangled with
such knotted structures so that it will become practically impossible to
separate \ these two types of polymers).To determine the recurrence/transience
for the complement of the trefoil knot requires some additional efforts which
are worth spending,see the next subsection.Already now few remarks are in
place. Just by looking at Fig.10 one can make the following observations. If
we collapse the solid line loops to lines and the dashed triangles to points,
we would obtain three-valent Bethe lattice as depicted in Fig.9. This lattice
is the universal covering space for the triangle depicted in Fig.11.%
\begin{figure}
[ptbptbptbptbptb]
\begin{center}
\includegraphics[
height=2.2857in,
width=2.5789in
]%
{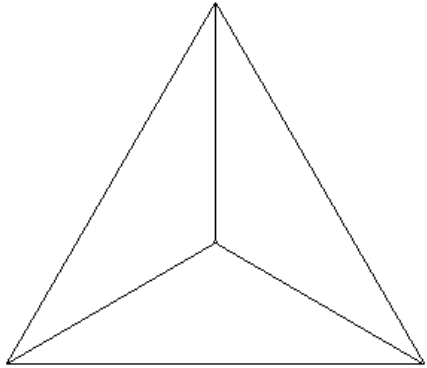}%
\caption{Base space for the graph depicted in Fig.10}%
\end{center}
\end{figure}
Actually, to be consistent with the existing conventions of combinatorial
group theory[83] one has to modify Fig.11 as depicted in Fig.12.%
\begin{figure}
[ptbptbptbptbptbptb]
\begin{center}
\includegraphics[
height=2.6991in,
width=3.0338in
]%
{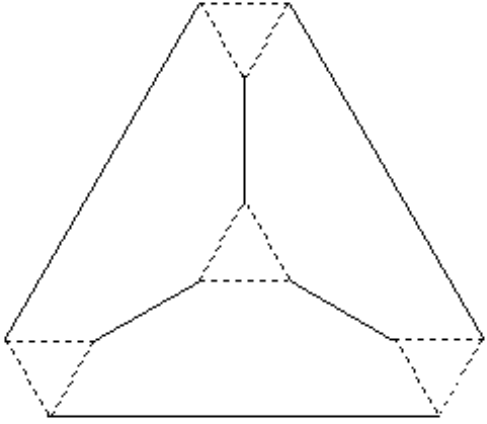}%
\caption{The same base space as in Fig.11 but with better resolution required
by the graphical representation of groups[]}%
\end{center}
\end{figure}
This is needed because ,according to the existing rules , at given vertex only
two edges of the same color(e.g. solid lines and/or dashed lines) can meet.By
looking at Fig.10 this fact then explains why figure 8 complement is transient
while that for the trefoil is recurrent. Indeed,for the figure 8 we have at
each vertex the probability 3/4 to go forward and only 1/4 to go backward
while for the trefoil we get 1/2 to go forward and 1/2 to go backward ,much
like for the random walks on the line which represent the universal covering
space for once punctured plane discussed in section 2. Surely,such walks are
recurrent as it had been demonstrated already.

Since the entanglements affect both static and dynamic properties of polymer
solutions,it is clear, that the presence of knotted structures should be
important in changing these \ properties. The presence of knotted structures
is also important biologially since for \textbf{different} knots the first
return and the escape times should be different .If the diffusion processes
take place in the vicinity of such knotted structures,e.g. DNA's, the outcomes
will be different for different types of knots.This observation could be
important in processes which involve the molecular recognition. Let us however
return to the problem of transience/recurrence for the random walks on
fundamental groups for thefoil \ and figure 8 in order to discuss additional
very important aspects of the recurrence-transience problem.

\subsection{Classical and quantum chaos and random walks on Teicm\"{u}ller
modular group}

From the knot theory textbooks,e.g.see Ref[43,81], it is well known that
although figure 8 and trefoil knots look very similar, they actually belong to
completely different ''universality classes'' as it can be seen already from
the difference in their group presentations and will be explained further
below. The trefoil knot is just a typical representative of the class of torus
knots while the figure 8 knot belongs to the class of hyperbolic knots
(actually it is the simplest example of hyperbolic-type knots[84]). It is not
our purpose to go into all details of the above distinction between these
different classes of knots. Instead, we are going to emphasize aspects of
these knots which are logically related to the discussions we had presented so far.

In our earlier work, Ref[1], we had noticed the fact that the fundamental
group of the the the punctured torus is the same as for figure 8.This could be
easily seen from the following picture, Fig.13.%
\begin{figure}
[ptb]
\begin{center}
\includegraphics[
height=2.2866in,
width=2.1862in
]%
{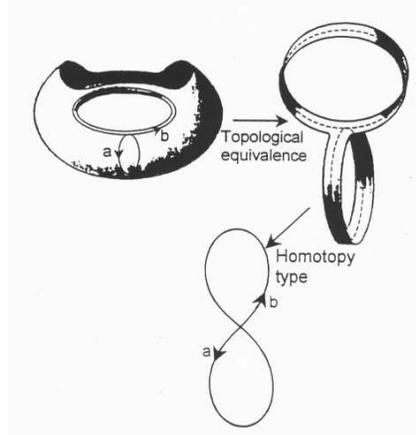}%
\caption{Topological structure of once punctured torus}%
\end{center}
\end{figure}
Although in physics literature quantum mechanics of such ''leaky torus'' had
been discussed [85] along with many generalizations[9-12], including those
which involve the quantum Hall effect[86],e.g.Aharonov-Bohm effect under the
periodic boundary conditions,etc.,surprisingly, the spectral analysis and
other related aspects of the punctured torus are still at the forefront of
research in mathematics[87].Because of this, we would like to emphasize here
some aspects of the problems related to punctured torus which had not been (to
our knowledge) discussed in physics literature.Following Ref.[88] ,it is
convenient to group the generators $a$ and $b$ of the free group $G_{8}$
according to the following scheme:
\[%
\begin{array}
[c]{ccc}%
ab & \rightarrow &  L\\
ba & \rightarrow &  L\\
(bb)^{-1} & \rightarrow &  L\\
ab^{-1} & \rightarrow &  R\\
bb & \rightarrow &  R\\
b^{-1}a & \rightarrow &  R
\end{array}
\]
where $R$ and $L$ stand for ''right'' and ''left'' as it is usually done in
symbolic dynamics[89].Surely, the above correspondence (automorphism) is quite
arbitrary and it is obvious that other possibilities could be tested. What is
less trivial is to prove[90] that the totality of words generated by sequences
given by Eq.(3.4) coincides exactly with the totality of words of the type
\begin{equation}
W=R^{n_{1}}L^{n_{2}}....... \tag{7.4}%
\end{equation}
where [n$_{1}$,n$_{2}$,...] is continuous fraction expansion of some number
,say $\theta<1.$ Recall [91], that the continuous fraction for $\theta$ is
actually given by
\begin{equation}
\theta=\frac{1}{n_{1}+\dfrac{1}{n_{2}+\dfrac{1}{n_{3}+...}}} \tag{7.5}%
\end{equation}
The continuous fraction always stops if $\theta$ is rational number or it
becomes periodic for some irrational numbers,e.g. those which come from
solutions of quadratic equations, and it is infinite and nonperiodic for
generic irrational $\theta$ . We would like now to argue that for the
torus-type knots $\theta$ is always rational number while for hyperbolic knots
,e.g. for figure 8, $\theta$ should be irrational . Since in the first case
words have finite length this means that if we have one of such longest words
,then multiplication by $R$ (or by $L)$ should make such word trivial, i.e.
W=1. The trivial word, when depicted graphically[83], simply means that there
is a closed loop in the graph. This,in turn, means that the motion is
recurrent .If there are no closed loops,then the motion could be either
transient or recurrent. . We had established this fact already in the previous
section for the figure 8 . Since the trefoil knot is associated with
three-valent Bethe lattice it means that the random walks on such lattices
behave quite differently as compared with Bethe lattices with vertices of even
valency . This difference also should show up in various properties of
statistical mechanical models on such graphs[80].In this work ,naturally, we
are not going to touch these subjects.Let us now explain better how all this
is associated with chaos. Superficially,an infinite word, Eq.(7.4), made of
totally random sequences of $R$ and $L$ already suggests chaoticity. In
general, the word W may or may not be random . It may correspond to the
sequence of base pairs along DNA, to some number written in binary system, the
sequence of ''up'' and ''down'' spins in Ising -like model, to sequence of
hits in the Poincare return maps, etc .One can do still better than just these
observations. To this purpose some knowledge of the knot theory is helpful and
we refer our readers to the specialized literature,e.g. see Refs[43,81].

Let is begin with a very well known fact from the combinatorial group theory
which is this. Let G be free nonabelian group made of two generators $a$ and
$b$. Let us construct a commutator $aba^{-1}b^{-1}\equiv\mathcal{K}$, then the
quotient $G$/$\mathcal{K}$ is just an abelian group that is , instead of
presentation given by Eq.(3.5), we obtain now
\begin{equation}
G_{a}=<a,b\left|  ab=ba>\right.  \tag{7.6}%
\end{equation}
and this is just the fundamental group of \ the torus. Such abelianization had
effectively removed the puncture from the torus so that it became essentially
the product of two circles with the group Z$\oplus Z$ of covering
transformations instead of just Z considered in section 2.This procedure of
abelianization is characteristic not only for the punctured torus but for
Riemann surface of \textbf{any }genus g :it is sufficient to make just one
hole in such surface in order to obtain the free group of 2g generators.Using
this fact ,one can prove the following related theorem

\textbf{Theorem} 7.1. \textit{\ If }$\Gamma$\textit{\ is finite graph with
}$\alpha_{1}$\textit{\ vertices and }$\alpha_{2\text{ }}edges$ \textit{then,
the fundamental group of the graph }$\pi_{1}(\Gamma)$\textit{\ is just a free
group of n generators where }$n=1+\alpha_{2}-\alpha_{1}.$

\textbf{Proof}.Please,consult Ref[43],page228.

The above theorem and the results presented in previous sections explain at
once the results of Ref[13] where the hyperbolicity of the disordered network
was established using much longer chain of arguments.

Going back to the torus case,one might think that the quantum mechanics on the
torus is trivial. To a large extent this is the case, as compared with the
case of the punctured torus, and ,indeed, many results for toroidal topology
had been obtained in the theory of dynamical systems[16],quantum billiards[92]
and the conformal field theories[93]. Nevertheless, even this case is still
under detailed investigation as recent literature indicates[94]. In any event,
by analogy with the circle, we shall utilize the \ properties of the covering
space which is just the infinite two dimensional lattice. We would like to
construct explicitly the group representation which describe the motions on
such lattice(s).Fortunately, this has been done already so that we can use
known results.Moreover,these results will be useful also for the punctured
torus case.Since the torus T=$S^{1}\times S^{1}$ and the case of S$^{1}$ was
studied in section 2 ,we can begin our discussion by dissecting our torus so
that it becomes a square(without loss of generality)with sides of unit length
. Then, the location of some point on the torus(or inside the square) can be
described in terms of the four real coordinates
\begin{equation}
(x_{1},x_{2},x_{3},x_{4})=(\cos\alpha,\sin\alpha,\cos\beta,\sin\beta)
\tag{7.7}%
\end{equation}
or,alternatively, in terms of two complex coordinates $z$ and $w$ such that
\begin{equation}
\left|  z\right|  ^{2}+\left|  w\right|  ^{2}=2 \tag{7.8}%
\end{equation}
since $\left|  z\right|  ^{2}=\left|  w\right|  ^{2}=1.$ But Eq.(7.8) is the
equation for 3-sphere
\begin{equation}
S^{3}=\{(z,w\}\in C^{2}:\left|  z\right|  ^{2}+\left|  w\right|  ^{2}=2\}
\tag{7.9}%
\end{equation}
It can be proven[95](and the proof is not too difficult) that all torus knots
are obtained as result of intersection of the sphere S$^{3}$ with the complex
surface $z^{p}+w^{q}$ =0 where p and q are relative primes so that p/q is some
rational number.Although all this is mathematically correct, we would like to
provide intuitively more convincing picture of what is actually taking place.
For this reason, we need to use the universal covering surface which is square
lattice with the side length being 2$\pi$ .For construction of the trefoil
knot, we need \ to consider the basic rectangle with sides of length 4$\pi$
and 6$\pi$ respectively which can cover the entire complex plane by obvious
translations. Let us now draw the straight line from the point (0,0) to the
point (4$\pi,6\pi)$ and let us by means of obvious translations bring the
squares affected by this line to the initial square which corner is located at
(0,0). All this is depicted on Fig.14.%
\begin{figure}
[ptb]
\begin{center}
\includegraphics[
height=3.8251in,
width=2.1197in
]%
{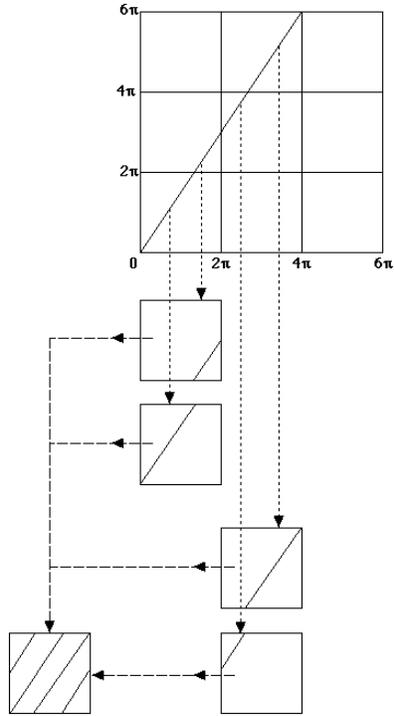}%
\caption{Dynamics of rational paths on the covering space of the torus}%
\end{center}
\end{figure}
Now,the pattern in this initial square can be transferred back from the
universal cover to the dissected torus without any change.Actually,this
pattern could be created directly in the base space too as depicted in Fig.15.%
\begin{figure}
[ptbptb]
\begin{center}
\includegraphics[
height=2.2866in,
width=2.2866in
]%
{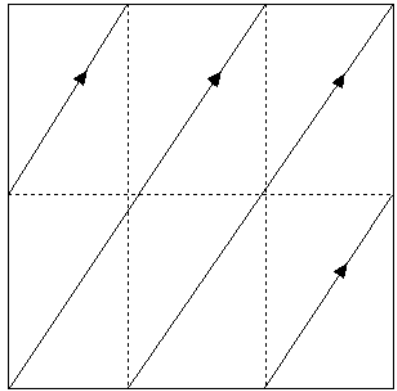}%
\caption{Dynamics of the rational paths in the base space of the torus}%
\end{center}
\end{figure}
This figure was \ created with account of the periodic boundary conditions on
the torus and,already from this picture one can recognize the knot projection
if one properly resolves the intersections between the dashed and the solid
lines(as it is always required in knot theory[43]).Fig.15 allows us to reglue
the square into the torus knowing that ,indeed,the three dimensional knot is
going to be created as result of such operation. This is depicted in Fig.16.%
\begin{figure}
[ptbptbptb]
\begin{center}
\includegraphics[
height=2.2857in,
width=3.0805in
]%
{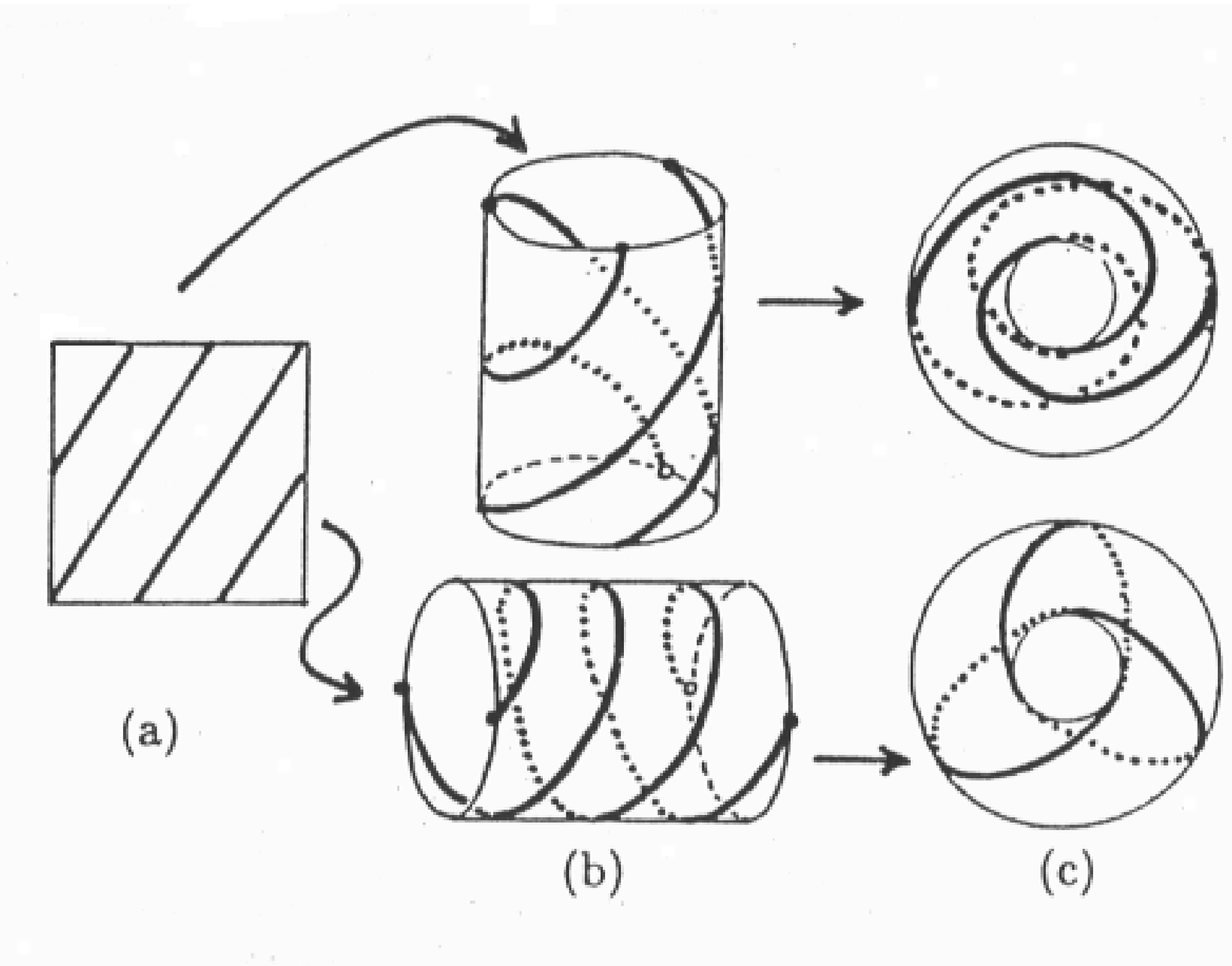}%
\caption{Rational dynamics on the torus can be equivalently described in terms
of the associated with it torus knots}%
\end{center}
\end{figure}
On this figure two trefoils are depicted :one is (3,2) and another is (2,3)
which are, actually, the mirror images of each other and are called the right
and\ the left hand trefoils respectively[43]. Now,what all this has to do with
our original intentions?.For this,we have to talk about the analytical theory
of toral automorphisms.It is clear,even without any actual calculations,that
the figure 8 knot should be associated with the lines of irrational slope in
the covering lattice, Fig.14.Such plausible reasoning would be to a large
extent correct if we would forget the results of section 3 in which we had
established that the universal covering space of the punctured torus is
\textbf{not} a square lattice but rather the Poincare upper half plane
H$^{2}.$ This fact provides yet another evidence of profound distinction
between the trefoil and figure 8 knots.The situation can be repaired somehow
if we provide different interpretation of the results depicted in Fig.14. \ To
this purpose one needs to introduce the notion of the Teichm\"{u}ller space of
the punctured torus[96]. To make this concept readily comprehensible let us
start with the notion of Dehn twists.As it had been proven by Nielsen[74],
following works of Dehn[82],all selfhomeomorphisms of two dimensional surfaces
can be performed with help of sequence of Dehn twists. Fig.17a depicts one of
such twists, while Fig.17b demonstrates what such twist can produce on the
surface of a torus.%
\begin{figure}
[ptbptbptbptb]
\begin{center}
\includegraphics[
height=2.0807in,
width=5.9767in
]%
{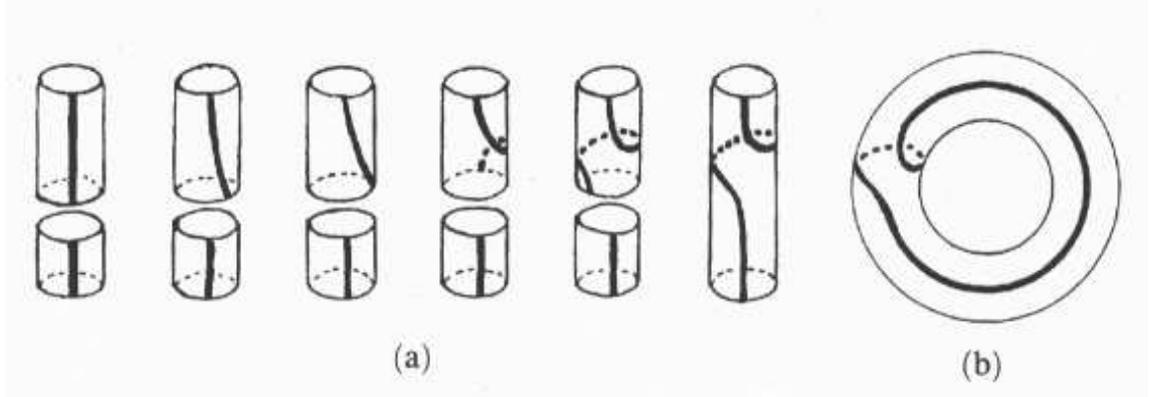}%
\caption{Physics of the Dehn twist}%
\end{center}
\end{figure}
This simple observation can now be broadly generalized. To do this several
steps are required. To begin, let us consider ,instead of a square lattice,
Fig.14,more general lattice ,as depicted in Fig.18,%
\begin{figure}
[ptbptbptbptbptb]
\begin{center}
\includegraphics[
height=2.6074in,
width=3.5146in
]%
{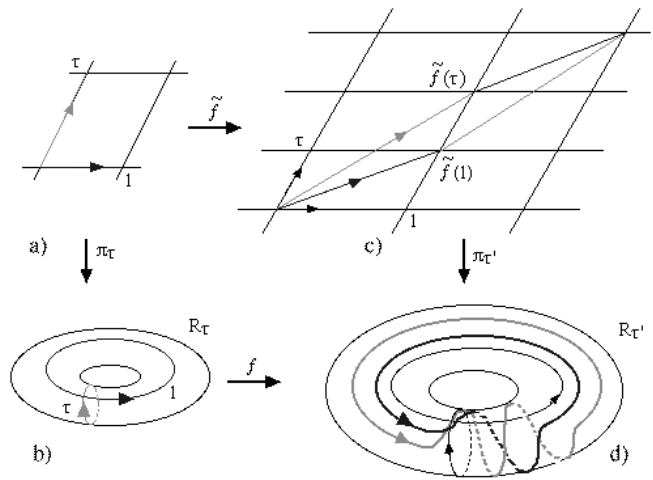}%
\caption{Canonical marking and dynamics of toral automorphisms}%
\end{center}
\end{figure}
which is characterized by the complex parameter $\tau($which is the ratio of
periods $\omega_{1}$ and $\omega_{2}$ (some relatively prime integers)along
the respective axes) so that ,in accord with Eq.(7.6), the abelian group of
translations $\Gamma_{\tau}$ can be described as
\begin{equation}
\Gamma_{\tau}=\{\gamma=m+n\tau\left|  m,n\in Z;\tau\in C,\operatorname{Im}%
\tau>0\}\right.  \tag{7.10}%
\end{equation}
Surely,the difference between different tori lies in different values of
$\tau.$ Let us reglue now the basic parallelogram in order to form torus with
canonical marking,Fig.18b. For the reference torus we can take,say $\tau
=i.$Now,if we change $\tau$ we thus going to form new torus which ,evidently,
\textbf{cannot} be \textbf{isometrically} transformed to the old one. The
relation between the new torus and the old one is established through
\begin{equation}
\tau^{^{\prime}}=\frac{a\tau+b}{c\tau+d} \tag{7.11}%
\end{equation}
with $a,b,c$ and $d$ being integers subject to constraint $ad-bc=1$.The group
PSL(2,Z) of the transformations just described is evidently the subgroup of
PSL(2,R) introduced in section 3 and is known as \textbf{modular }group.Since
PSL(2,R) is the group of isometries of H$^{2}$ ,evidently, PSL(2,Z) is also
acting on H$^{2}$. The moduli space M=H$^{2}/PSL(2,Z)$ for the torus, by
definition ,coincides with the fundamental domain , Fig.19,%
\begin{figure}
[ptbptbptbptbptbptb]
\begin{center}
\includegraphics[
height=2.2866in,
width=2.2113in
]%
{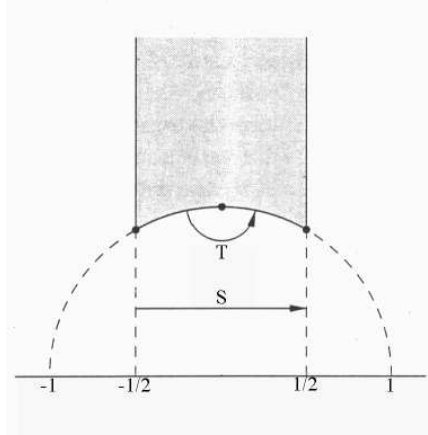}%
\caption{Fundamental domain for PSL(2,Z)}%
\end{center}
\end{figure}
for the group PSL(2,Z) as is well known[51]. The Teichm\"{u}ller space is
related to M but is more physically appealing as it will become clear
momentarily.To this purpose,let us discuss the sequence of transformations as
depicted in Fig.18c and d.From this figure it is clear, that upon regluing of
the deformed parallelogram the pattern of canonical markings ,Fig.18b, had
undergone some changes and,in view of Fig.17, these changes could be
interpreted in terms of the sequence of Dehn twists made with respect to
canonical markings.Let us discuss an example of the trefoil knot. Let we have
originally the square lattice with sides of the basic square 1 and $i$
respectively. To construct a new lattice we use as a guide Fig.18. The
fundamental paralleogram of new periods can be chosen according to the
following set of equations written with respect to the ''old'' axes:%

\begin{equation}
\tilde{f}(\tau)=\omega_{1}\tau^{\prime}=ai+b \tag{7.12}%
\end{equation}
where in Eq.(7.12) we have to put $a=3$ and $b=2$ in order to obtain the
trefoil knot. The numbers 3 and 2 are precisely the numbers of full twists
with respect to canonical basis. In addition to Eq.(7.12) we also have%

\begin{equation}
\tilde{f}(1)=\omega_{1}=ci+d \tag{7.13}%
\end{equation}
and
\begin{equation}
ad-bc=1 \tag{7.14}%
\end{equation}
The above two additional equations reflect the fact that the transformation
leading from the original square to the final parallelogram are
area-preserving which is insured by Eq.(7.14) while Eq.(7.13) comes as result
of such constraint. With $a=3$ and $b=2$ Eqs. (7.14) produce the following set
of solutions for $c$ and $d$:a)c=1,d=1 or b)c=-2 and d=-1.Fig.s 20%
\begin{figure}
[ptb]
\begin{center}
\includegraphics[
height=2.2874in,
width=2.3091in
]%
{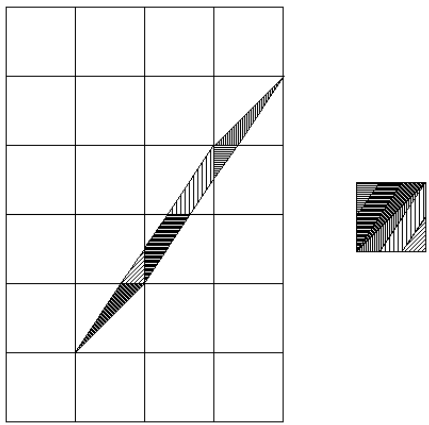}%
\caption{An example of an area preserving toral homeomorphism related to
formation of the trefoil knot}%
\end{center}
\end{figure}
and 21%
\begin{figure}
[ptbptb]
\begin{center}
\includegraphics[
height=2.2874in,
width=2.2373in
]%
{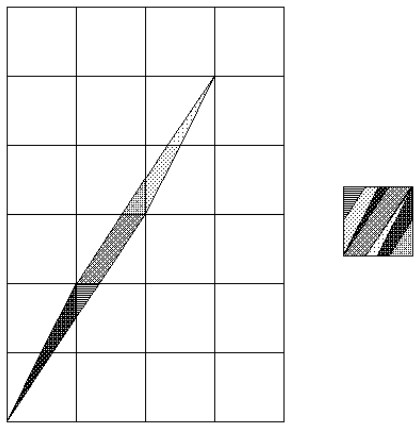}%
\caption{Different toral homeomorphism associated with the same trefoil knot}%
\end{center}
\end{figure}
demonstrate all this graphically. Both solutions make sense for the following
reasons. These solutions provide the number of Dehn twists for the second
basic curve involved in canonical marking. This second basic curve intersects
the first one in just one point ,Fig.18b,initially and Fig. 18d,finally.But
the amount of twisting of this second curve is related to the first one only
by the law of area conservation so that in both cases we get the trefoil knot
for the first basic curve.One can express all this in terms of vectors (a,b)
and (c,d) which provide the numbers of Dehn twists along ''vertical''
/''horizontal'' directions for the first and second canonical curves
respectively.After this discussion it becomes clear that if we reglue the
parallelograms in Fig.20 and 21, we reobtain respective toruses as depicted in
Fig.18d. Evidently, if we \textbf{physically} relate the initial square to the
final parallelogram(s) we recognize that some stretching (may be twisting
also)was necessary in order to bring the square to the parallelogram before
gluing. Hence, the Dehn twists are \textbf{always} associated with some
stretching and twisting of the underlying surface. This fact is
\textbf{directly} associated with the notion of the \textbf{Teichm\"{u}ller
space} since it is just the space of parameters responsible for surface
deformations. In our case,the driving parameter is $\tau$ so that the
Teicm\"{u}ller space is H$^{2}$ for both torus and punctured torus[96] as it
should become now intuitively clear based on the discussion we had just
completed.Finally, the sequence of Dehn twists associated with
selfhomeomorphisms(diffeomorphisms) of torus(surface of genus g, in general)
causes some fundamental changes in the marking responsible for the fact
that,for example,different torus knots are \textbf{isometrically}
different.Therefore, it is possible to associate the \textbf{mapping class}
\textbf{group}, known also as \textbf{Teichm\"{u}ller modular group}, with the
group generated by all nontrivial (that is not reducible to identity) Dehn
twists (associated with canonical marking) so that the random walk problem of
section 6 becomes the problem about the random walk on the mapping class group
(except ,now we have to deal with random walks on groups which are not
free).An excellent and readable description of the mapping class groups could
be found in Ref.[97] while good discussion of random walks on various groups
which are not free could be found in Ref[21].Our first illustration of
nontriviality of this group is the isometric difference between the tori
depicted in Figs.20 and 21. Let us discuss this topic a bit further since it
has some applications to dynamics of liquid crystals and 2+1 quantum gravity
as described in our earlier work,Ref[17],(and ,especially, Appendix of this
reference where connections of these dynamical problems with toral
automorphisms (in the simplest case)are discussed) .

If we select $\tau$ in the fundamental domain,Fig.19, i.e. in the moduli space
M, then, it can be shown[51] ,that different $\tau\in M$ are \textbf{not}
connected with each other by means of M\"{o}bius-type
transformation,Eq.(7.11).But,once we picked some specific $\tau$ ,then, this
$\tau$ is connected with other $\tau^{\prime}$ which lies \textbf{outside} the
moduli space via transformation of the type given by Eq.(7.11) . By means of
successive applications of transformation law Eq.,(7.11), one can reach all
conjugate points in the corresponding hyperbolic triangles which tessellate
H$^{2}$ so that application of Eq.(7.11) to all points of the moduli space
will cover the entire H$^{2}.$ Thus the Teicm\"{u}ller modular group is acting
on the moduli space. Moreover, it is possible to show[51] that the arbitrary
matrix
\begin{equation}
V=\left(
\begin{array}
[c]{cc}%
a & b\\
c & d
\end{array}
\right)  =TS^{n_{1}}S^{n_{2}}T\cdot\cdot\cdot S^{n_{k}}TS^{n_{l}} \tag{7.15}%
\end{equation}
where
\begin{equation}
S=\left(
\begin{array}
[c]{cc}%
1 & 1\\
0 & 1
\end{array}
\right)  \tag{7.16}%
\end{equation}
and
\begin{equation}
T=\left(
\begin{array}
[c]{cc}%
0 & 1\\
-1 & 0
\end{array}
\right)  \tag{7.17}%
\end{equation}
Since
\begin{equation}
T^{2}=\left(
\begin{array}
[c]{cc}%
-1 & 0\\
0 & -1
\end{array}
\right)  =\left(
\begin{array}
[c]{cc}%
1 & 0\\
0 & 1
\end{array}
\right)  \tag{7.18}%
\end{equation}
projectively,e.g.see Eq.(7.11), this explains why in expansion (7.15) all T's
have the first power. The fundamental domain for PSL(2,Z) in Fig.19 is
depicted with \ generators T and S in mind. As for the generator S ,we had
already encountered it ,e.g.see Eq.(3.6). This means :a) that the numbers
n$_{1},....,$n$_{l}$ can be arbitrary integers and b) that using
Eqs(7.15)-(7.17) we can always replace a -generator in Eq.(3.6) with the
appropriate sequence of T and S generators. Moreover,the combination $TS=Q$
produces projectively
\begin{equation}
Q^{3}=1 \tag{7.19}%
\end{equation}
so that,in view of Eq.s(7.3) ,(7.18) and (7.19), we can consider the random
walk on the complement of the trefoil using the generators of PSL(2,Z). This
result is in accord with Ref[81] where other arguments were used to reach the
same conclusions. From the discussion we had just presented it is clear that
the description of the toral automorphisms by means of the random walks on the
mapping class group should be equivalent to traditionally used in the theory
of dynamical systems [16].E.g.,for example,the torus mapping depicted in
Fig.20 could be equivalently described in terms of the map given by
\begin{equation}
y^{\prime}=x+2y \tag{7.20a}%
\end{equation}%
\begin{equation}
x^{\prime}=x+3y \tag{7.20b}%
\end{equation}
and,of course,
\[
1\cdot3-1\cdot2=1
\]
The above arguments (x and y) in this map should be taken mod1[16] by analogy
with S$^{1}$ case earlier discussed in section 2.For the case of toral
automorphisms the use of Teicm\"{u}ller space interpretation appears as
plausible curiosity. The things change dramatically if the self homeomorphisms
of surfaces of higher genus is of interest, e.g. in gravity or liquid
crystals[17]. In mathematics literature ,the Nielsen-Thurston theory of
surface selfhomeomorphisms is well developed[24,98] and,in view of the
discussion already presented, can be used for description of various physical
phenomena, including dynamics of liquid crystals and gravity .Indeed,based on
the results just obtained the toral selfhomeomorphisms could be of three
types:a)\textbf{reducible},when the words like that in Eq.(7.15) are of finite
length (this corresponds to finite continued fraction
expansion,Eq.(7.5));b)\textbf{periodic(}which may correspond to the periodic
continued fraction of some quadratic irrationals)and ,c)\textbf{irreducible}
(or Anosov (pseudo Anosov in more general case))which may be associated with
truly irrational continued fraction expansions. The first possibility in the
language of liquid crystals[99] corresponds to solid phase,the second to the
hexatic phase and the third to the liquid phase. For more details,please,
consult our earlier work,Ref[17].

\subsection{\bigskip Three-manifolds which fiber over the circle}

Connections between theory of dynamical systems and theory of \ knots and
links are the most naturally described in terms of the Nielsen-Thurston
theory[98] and ,although this field \ is still under active development even
in mathematics[100] , we would like to provide here a sketch of what one may
expect upon development of such connections.This information may be useful in
fields ranging from gravity through liquid crystals, dynamics of fracture and
even folding of proteins(for references related to these problems
,please,consult our earlier work,Ref[17]).

Description of dynamical systems in terms of Lyapunov exponents, fractal and
multifractal dimensions,entropy, etc is well known[101]. Recently, however,
more and more emphasis is being made on connections between dynamical systems
and the theory of knots and links[102] . The complements of knots and links
are 3-manifolds and,moreover, there are 3- manifolds which are not related to
links /knots ,especially those which are not compact.The 3-manifolds arise
completely naturally in the theory of dynamical systems. To realize this fact
it is sufficient just to take a look again at Fig.15. Extension this picture
to a cube with periodic boundary condition creates already our first 3
manifold, i.e. cube with the opposite sides properly identified. This example
already provides a clue as to how 3-manifolds can be constructed: e.g. take
some polyhedron and identify the opposite sides in some systematic way. As a
result of such an identification we again obtain some 3-manifold. Since some
of these 3 manifolds are related to knots it is interesting to know that, for
example, Fig.8 knot complement ,which is also a 3-manifold, can be obtained by
proper gluing of faces of two ideal tetrahedra to each other. This and other
very beautiful \ illustrations of making 3-manifolds associated with simple
knots/links could be found in Ref.[103]. In case of trefoil ,the associated
3-manifold is a (3.2) Lens space. It is easily created out of two
\textbf{solid} toruses.To do so we should have one torus with standard
canonical marking being glued to other (twisted) torus which contains the
trefoil knot on its surface so that the initial (canonical) marking is glued
to the final marking and then the rest of two torus surfaces being glued to
each other [104]. This example gives an idea about the way 3-manifolds are
related to surface dynamics.Let us now be more specific. Let $S$ be some
surface and let x$\in S$ .Consider a homeomorphism $\psi:S\rightarrow S$ .
Imagine now that this homeomorphism takes place in stages ,i.e.evolvs in time
so that at time t=0 $\psi(x)=x$ while at time 1 $\psi(x)$ may or may not
coincide with x $\forall x\in S.$ The time 1 is, of course,quite arbitrary and
is chosen for convenience only.Let us now identify (x,0) with ($\psi(x),1)$
$\forall x\in S.$This identification leads us to a \ 3-manifold T$_{\psi}$ .
This 3-manifold is a fiber bundle with fiber S, base S$^{1}($a circle) and
monodromy $\psi$. It is clear that the monodromy $\psi$ is directly related to
the random walk on the Teichm\"{u}ller modular group (as described in the
previous subsection).It is clear also that different surface dynamics will
lead to different 3-manifolds in general so that the topology and geometry of
3-manifolds is yet another way of description of dynamical systems[100,102].

The above picture can be elaborated now as follows.Since there is one-to-one
correspondence between the sequence of Dehn twists and the sequence of words
being regulated by the continued fractions expansions it is clear that the
trivial fiber bundle can be associated with the \textbf{reducible}
selfhomeomorphisms and that the \textbf{periodic }selfhomeomorphisms produce
naturally 3-manifolds known also as Seifert fibered spaces[81,84,104]. Surely,
all torus knots are associated with periodic selfhomeomorphims and,therefore
are in one- to-one correspondence with Sefert fibered spaces as was already
established by Sefert in 1933(the sufficiently simple proof of this could be
found in chr.6 of Ref.[81]).What is much less clear why the above construction
of 3-manifolds should be also applicable to \textbf{irreducible (}or pseudo
Anosov\textbf{) }selfhomeomorphisms. The fact that this is actually possible
had been proven rather recently [100].And ,the most remarkable fact of this
proof lies in use of Feigenbaum-Cvitanovich results for one dimensional maps
which, in the light of the results we had presented already should not be
totally unexpected. 3-manifolds associated with knots such as figure 8 are
hyperbolic. So that all knots other than torus knots should be associated with
hyperbolic 3-manifolds. That this is indeed the case had been proven by
Thurston[86,103] and was mentioned already in our earlier work, Ref[14]. The
hyperbolicity of knots is directly connected with transience of random walks
in the presence of such knots as we had already discussed in section 7.A.
\ Moreover, using the results of hyperbolic geometry it is possible to get
\ information about fractal dimensions of the limits sets (in case of
dynamical interpretation). Some introduction to this subject can be found in
our earlier work,Ref.[105].

\subsection{Schottky doubles, Markov triples \ and Grothendieck's dessins d'enfants}

Although we had covered many aspects of the figure 8 story, there are many
topics which we had not mentioned at all so far. In this subsection we would
like to correct this deficiency. Evidently,our presentation is going to be
brief but, since the topics to be discussed are directly and logically related
to the rest of this paper, it is impossible not to mention about them .

Let us begin with the notion of Schottky double.From the discussion we had so
far it is clear that treatment of boundaries in the hyperbolic geometry is not
as simple as it is in Euclidean case. For instance, suppose we would like to
calculate the spectrum of Laplacian on the punctured torus .The question
immediately arises: how the spectrum of Laplacian will depend upon the
boundary conditions at the puncture ? To some extent this problem was solved
by Gutzwiller[85] who however admits,e.g. see page 376 of Ref.[85], that more
systematic mathematical treatment is given in the monograph by Lax and
Phillips[106]. Surprisingly, the case is still not closed as recent Annals of
Mathematics paper [87] indicates. Hence, it would be too naive to address this
problem in its full generality in this paper. Nevertheless, we would like to
mention one simple case which has its familiar analogue,for example, in
standard quantum mechanics. The idea belongs to Schottky and is described in
Ref.[96].Consider a Riemann surface R \ with punctures and its copy \={R}
obtained by reflecting R in the mirror. Given these two copies , glue them
along the boundaries of punctures . Thus obtained new Riemann surface is known
as Schottky double. In the case of punctured torus the Schottky double is a
double torus which ,not surprisingly, was also considered by Gutzwiller[85].
Such construction is equivalent of having perfectly reflecting boundary
conditions in standard quantum mechanics and also has some applications to
polymer solutions in confined geometries[14] ,etc.The boundary conditions(that
is the existence of puncture), as pointed by Gutzwiller [85] and known from
the theory of Riemann surfaces[107] ,could be modelled by imposing certain
constraints on the commutator $\mathcal{K}$ defined before Eq.(7.6).In
particular, one can require
\begin{equation}
\mathcal{K=}\mathit{aba}^{-1}b^{-1}=\mathcal{P} \tag{7.21}%
\end{equation}
(where $\mathcal{P}$ is \ transformation of the type given by Eq.(7.11)) to be
\textbf{parabolic }, that is tr($\mathcal{P}$)=2. Such requirement provides
one constraint on the explicit form of matrices $a$ and $b$.Additional
constraints could be imposed by locating vertices of the hyperbolic triangle
at points 0, 1 and $\infty$ as depicted in Fig.2, or hyperbolic quadrangle at
points -1,0,1 and +$\infty$ , as depicted in Fig.3. Let us check this result
explicitly.Using results of section 3 we obtain,
\begin{equation}
a(z)=\frac{z}{2z+1} \tag{7.22}%
\end{equation}%
\begin{equation}
b(z)=z+2 \tag{7.23}%
\end{equation}
Accordingly,
\begin{equation}
a(z)^{-1}=\frac{z}{-2z+1} \tag{7.24}%
\end{equation}
and
\begin{equation}
b(z)^{-1}=z-2 \tag{7.25}%
\end{equation}
Using Eqs.(7.22)-(7.25) in (7.21) a short computation produces
\begin{equation}
\mathcal{P(}z)=\frac{\frac{3}{8}z+1}{\frac{1}{2}z+\frac{13}{8}} \tag{7.26}%
\end{equation}
From here,tr($\mathcal{P(}z))$=2 as required. The above simple calculation
teaches us some very important lessons.First , since we are dealing with
projective transformations ,e.g. that given by Eq.(7.11), we can always
normalize our matrices the way which is the most convenient for us. In
particular, already in more than hundred years ago Fricke[108] had shown that
for any unimodular matrices A and B
\begin{equation}
\left(  trA\right)  ^{2}+\left(  trB\right)  ^{2}+\left(  trAB\right)
^{2}=\left(  trA\right)  \left(  trB\right)  \left(  trAB\right)
+tr\mathcal{K}\text{+2} \tag{7.27}%
\end{equation}
where,as before,$\mathcal{K}$ is the commutator of A and B. Now , let
tr$\mathcal{K=-}$2.This is quite permissible in view of Eq.(7.26) since
multiplication by -1 of both numerator and denominator changes nothing(e.g.see
Eq.(7.18)).Such normalization however, makes a difference in Eq.(7.27) which
now could be brought to form
\begin{equation}
\left(  trA\right)  ^{2}+\left(  trB\right)  ^{2}+\left(  trC\right)
^{2}=\left(  trA\right)  \left(  trB\right)  \left(  trC\right)  \tag{7.28}%
\end{equation}
where C= (AB)$^{-1}$ and the fact that tr(D)=tr(D$^{-1})$ for any unimodular
matrix D was used.By setting $\left|  trA\right|  $=3$m_{1},\left|
trB\right|  $=3$m_{2}$ and $\left|  trC\right|  $=3$m_{3}$ we obtain ,instead
of Eq.(7.28),the following equation for the \textbf{Markov triples[}%
25\textbf{]}
\begin{equation}
m_{1}+m_{2}+m_{3}=3m_{1}m_{2}m_{3} \tag{7.29}%
\end{equation}
with $m_{1},m_{2},m_{3}$ being nonnegative integers.By looking at this
equation we immediately recognize that it is trivially satisfied by
$m_{1}=m_{2}=m_{3}=1.$ This information ,when combined with the requirement of
unimodularity, is sufficient for restoring the explicit form of matrices A and
B ,and ,hence, AB. Questions arise :a)are there other solutions to the
equation for Markov triples and b) what is their physical significance if they
indeed exist? The answer to the first question is known in mathematical
literature and will be discussed shortly below. The answer to the second
question could be related to the fractional quantum Hall effect. Indeed, if
solutions other than trivial to Eq.(7.29) do exist, ,then, based on the
results of sections 3 and 4, we conclude that they may be associated with
different shapes of the fundamental domains in H$^{2}$ plane and
,accordingly,to different locations of punctures in S$^{2}.$Thus if the
punctures are associated with anyons, then , evidently, their mutual position
cannot be arbitrary so that we have some sort of solid phase in the
\textbf{cluster approximation} (made of 3 punctures). Within this
approximation the filling fraction[14] ,0%
$<$%
$\nu<1$ can be calculated and, because the pattern or Markov triples is very
intricate (see discussion below) one obtains a very rich array of the
permissible filling fractions characteristic for the fractional quantum Hall
effect[109].To what extent such cluster picture can be extended to more than 3
punctures is related to the works of Grotendieck on dessin d'enfants[22] to be
also discussed below.

Looking at Eq.(7.29) it is clear that there is a complete symmetry between
$m_{1},m_{2}$ and $m_{3}.$ This symmetry can be broken if we notice that
(2,1,1) is also solution to Eq.(7.29). This \ solution is called
\textit{singular}[25].It could be shown ,that this is the only solution for
which 2 out of three Markov numbers are the same. All other solutions have
distinct Markov numbers.To generate these other solutions (Markov triples) it
is sufficient to have a ''seed'' \ in the form ($m_{1},m_{2},m_{3})$ which ,by
definition, obeys Markov equation, Eq.(7.29). Then, the first generation of
new Markov triples given by
\[
(m_{1}^{^{\prime}},m_{2},m_{3}),(m_{1},m_{2}^{^{\prime}},m_{3})\text{ and
(}m_{1},m_{2},m_{3}^{^{\prime}})
\]
where
\begin{equation}
m_{1}^{^{\prime}}=3m_{2}m_{3}-m_{1};m_{2}^{^{\prime}}=3m_{1}m_{3}-m_{2}%
;m_{3}^{^{\prime}}=3m_{1}m_{2}-m_{3} \tag{7.30}%
\end{equation}
.If \ one starts with (1,1,1) then, the Markov tree (its part of course) is
depicted in Fig.22.%
\begin{figure}
[ptb]
\begin{center}
\includegraphics[
height=4.9493in,
width=4.088in
]%
{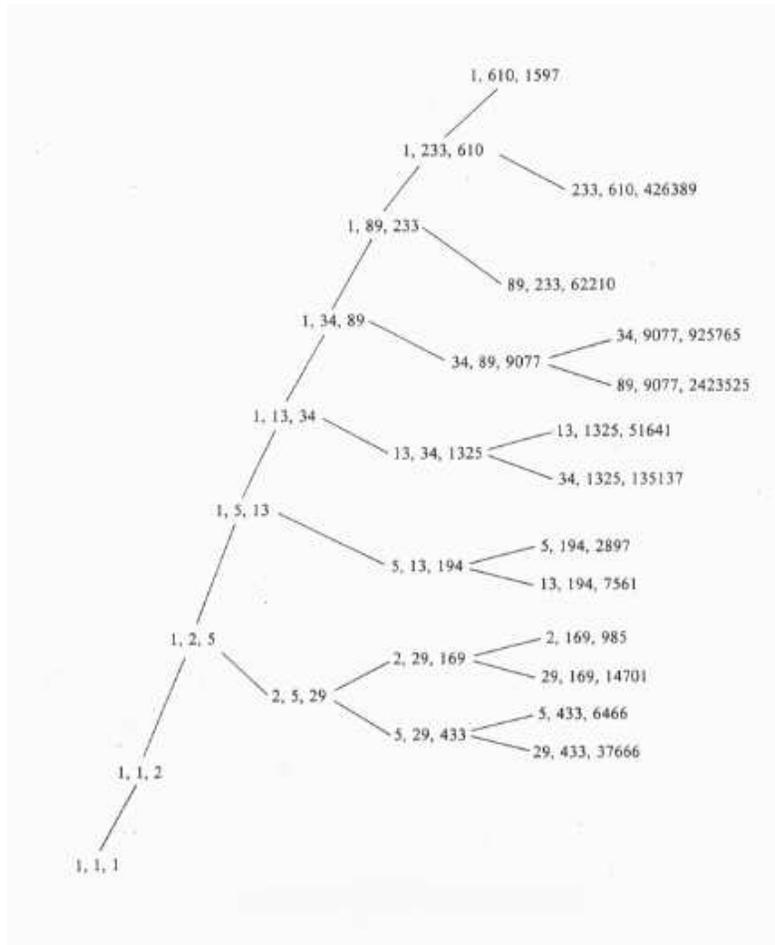}%
\caption{Fragment of the Markov tree closest to the original ''seed''}%
\end{center}
\end{figure}
The emergence of numbers $m_{1}^{^{\prime}}$ ,etc. , can be easily understood
if we consider ,instead of Eq.(7.29), its equivalent form given by
\begin{equation}
f(x)=x^{2}-3m_{2}m_{3}x+m_{2}^{2}+m_{3}^{2}. \tag{7.31}%
\end{equation}
Evidently, $f(x)=0$ for $x=m_{1}$ and $x^{^{\prime}}=3m_{2}m_{3}-m_{1}.$
Additional information about Markov triples could be found in Refs.[110-112].
Now the question arises as to what extent one can generalize these results to
more than tree punctures. There are several ways to proceed. For example,
prior to Theorem 7.1. we've noticed that it is sufficient to make one puncture
(hole) in surface of any genus g in order to obtain a free group with 2g
generators. By analogy with Eq.(7.29), one may think of solutions of equation
\begin{equation}
x_{1}^{2}+x_{2}^{2}+\cdot\cdot\cdot+x_{n}^{2}=nx_{1}x_{2}\cdot\cdot\cdot x_{n}
\tag{7.32}%
\end{equation}
as it was done by Hurwitz[113] and elaborated \ by Hirzebruch and Zagier[114].
There is ,however,much more striking approach to the whole problem . It was
proposed by Grotendieck [22] \ and was elaborated by Belyi[115] and
others[22].These mathematical results had found already their place in physics
literature in connection with problems related to discretized string
theory[22] and the theory of polygonal billiards[116].As it was argued above,
these results could be potentially useful in the theory of fractional Hall
effect and/or quantum chaos,etc.

The idea of the Grotendieck approach is rather simple. Take for example the
torus case. The torus ,Fig.18a(or 18b) can be made out of two triangles and
,accordingly, any Riemann surface of genus g also can be triangulated. Such
triangulation of arbitrary closed Riemann surface could be easily understood
if we would consider the motion of ficticious particle in a triangular
billiard whose angles are some rational fractions of $\pi.$ Instead of
watching the complicated trajectory inside the triangle it is much better to
unfold the trajectory in a way shown in Fig.23.%
\begin{figure}
[ptb]
\begin{center}
\includegraphics[
height=3.179in,
width=1.8386in
]%
{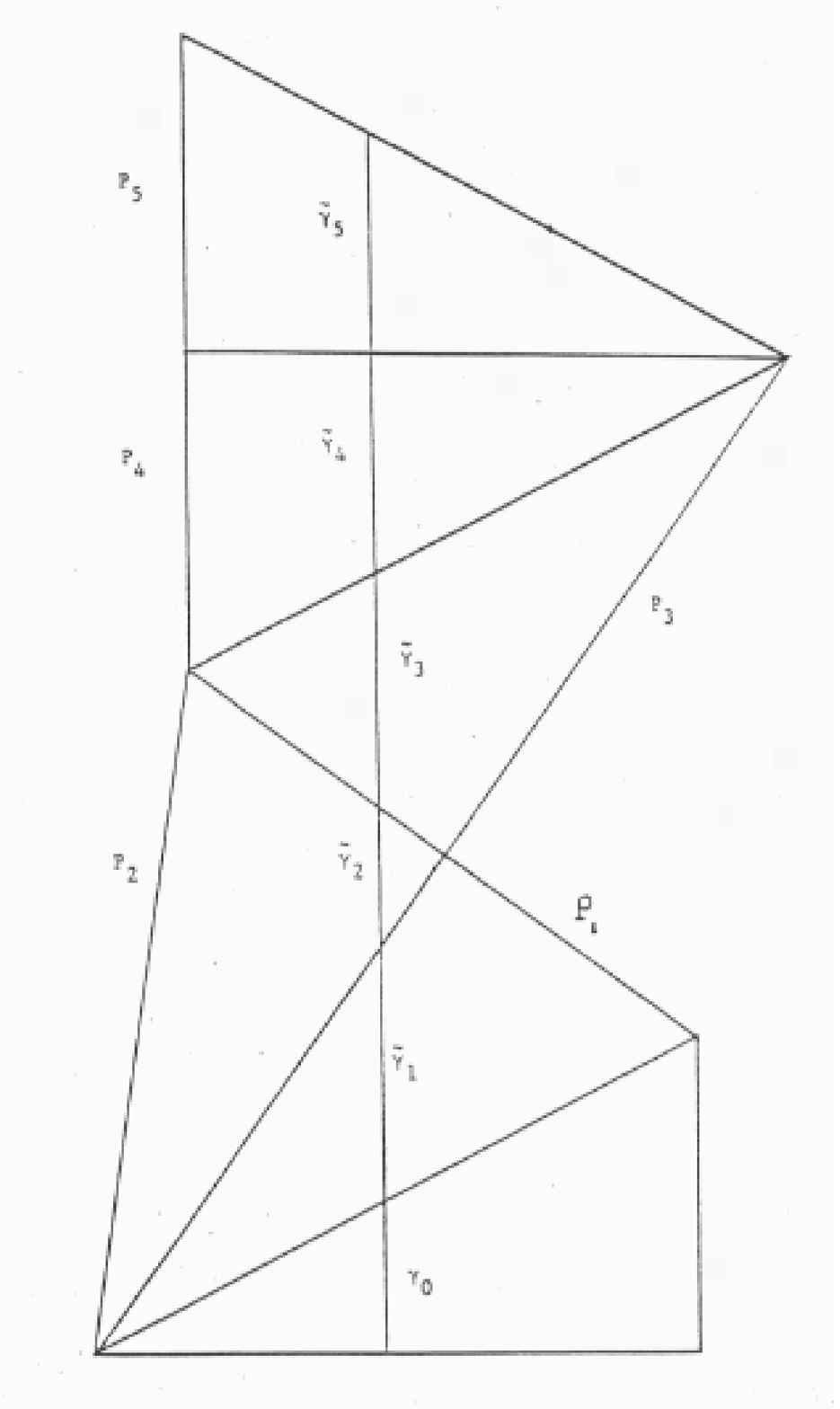}%
\caption{Unfolding of the particle trajectory in triangular billiards: each
time the particle hits the wall, one should extend the particle motion by
considering its motion in the mirror image of the triangle with respect to
this wall.}%
\end{center}
\end{figure}
Such unfolded trajectory is just a straight line ,just as in the case of a
torus ,Fig. 14. From the covering surface ,e.g. that depicted in Fig.14, one
needs now to restore the corresponding (Riemann) surface. This was done
initially by Richens and Berry[117] and was subsequently generalized and
elaborated by Zemlyakov and Katok[118] and many others,e.g.seeRef.[119].
Strictly speaking , to triangulate Riemann surface one needs to consider such
triangle groups whose action on the fundamental domain for such groups will
cover (tessellate) given Riemann surface without gaps or overlaps. Grotendieck
calls such groups ''cartographic groups''. A good survey of the cartographic
groups is given in Ref.[22].The triangulation of Riemann surfaces is not a new
idea.The new idea lies in utilizing the results for the trice punctured
sphere,sections 3-4, in order to achieve the triangulation of Riemann surface
of \textbf{any} genus.The dessins d'enfants (the drawings of \ children)
problem is directly related to the triangulation problem and can be formulated
as problem of drawing an arbitrary(but pre assigned) closed graph on Riemann
surface. This happens to be not a simple task .To make a connection between
the trice punctured sphere and some Riemann \ surface one needs to have a
mapping function which connects the triangulation on the Riemann surface with
triangles on the sphere.This function had been constructed explicitly by
Belyi[115] . We follow ,however, the work of Shabat and Voevodsky[120] in
order to illustrate the basic idea of the method.

If the surface is orientable, it is possible to color the adjacent triangles
in two different colors,say, black and white. Given such coloring, one can
distinguish three types of vertices :$\circ$ ,$\cdot$ and $\star.$Let the red
arrow connect $\circ$ and$\star$, let also the black arrow connect $\star$
with $\cdot$ and, finally, let the blue arrow connect $\cdot$ with $\circ$
.Let us next associate the vertex $\circ$ with number 0, $\star$ with 1 and
$\cdot$ with $\infty$ . The above numbers are associated with the location of
punctures on S$^{2}$ for the trice punctured sphere as discussed in sections 3
and 4.To understand the connection, observation of Fig.24a is helpful at this
time.%
\begin{figure}
[ptb]
\begin{center}
\includegraphics[
height=2.2866in,
width=3.6106in
]%
{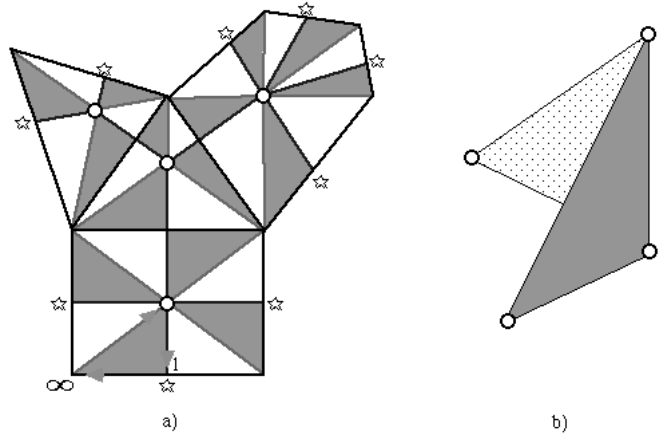}%
\caption{Triangulation of the arbitrary Riemann surface a) and the associated
with it double triangle b) which (upon regluing) is topologically equivalent
to the trice punctured sphere as discussed in the text}%
\end{center}
\end{figure}
Let us select two adjacent triangles which ,by construction, are colored
differently, e.g. see Fig.24b. Finally, if we glue together these triangles
along the free edges, thus obtained object is just a trice punctured
sphere.This can be easily understood if we imagine that the triangles are made
of rubber and we blow up air through one of the open holes and then close it.
Hence, to get a triangulation of an arbitrary Riemann surface $\mathcal{R}$ ,
one shoold only look for the appropriate branched covering with branches over
the points 0, 1 and $\infty.$ Clearly, for each point, say, 0 on S$^{2},$
there is a finite number of the corresponding points on $\mathcal{R}$ , etc.
The mapping function $\mathit{F(z)}$ which actually maps S$^{2}\backslash
\{0,1,\infty\}$ to $\mathcal{R}$ is given by
\begin{equation}
F(z)=\frac{\left(  m+n\right)  ^{m+n}}{m^{m}n^{n}}z^{m}(1-z)^{n} \tag{7.33}%
\end{equation}
with m, n being some integers which value depends upon the genus of
$\mathcal{R}$ .

From this discussion it is clear that thus obtained Markov triples could \ be
now used for an arbitrary $\mathcal{R}$ and this fact then proves their
relevance for computation of the multitude of the filling fractions $\nu$
characteristic for the factional Hall effect[109]. To complete our discussion
, we have to consider still one more topic.

\subsection{Connections with braid groups}

We include this subsection in this paper in order to emphasize some aspects of
theory of braids which are directly related to the discussion we had so far.
We assume ,that the reader is familiar with some basic concepts of the theory
of braids [81].The braid group is infinite group of n-1 generators $\sigma
_{1},\sigma_{2},...\sigma_{n-1}$ obeying the following set of defining
equations(in the planar case only)
\begin{equation}
\sigma_{i}\sigma_{j}=\sigma_{j}\sigma_{i}\text{ for }\left|  \text{i-j}%
\right|  \geq2 \tag{7.34}%
\end{equation}
and
\begin{equation}
\sigma_{i}\sigma_{i+1}\sigma_{i}=\sigma_{i+1}\sigma_{i}\sigma_{i+1}\text{ for
1}\leq i\leq n-2. \tag{7.35}%
\end{equation}
As it was shown already by Artin(inventor of the Braid group) [121] , the
braid group can actually be generated just by its two elements
\begin{equation}
a=\sigma_{1}\sigma_{2}...\sigma_{n-1} \tag{7.36}%
\end{equation}
and
\begin{equation}
\sigma=\sigma_{1}. \tag{7.37}%
\end{equation}
Thus,
\begin{equation}
\sigma_{i}=a^{i-1}\sigma a^{-(i-1)}, \tag{7.38}%
\end{equation}
and
\begin{equation}
a^{n}=(a\sigma)^{n-1}. \tag{7.39}%
\end{equation}
In particular, let n=3,then, instead of Eqs(7.35), we get
\begin{equation}
\sigma_{1}\sigma_{2}\sigma_{1}=\sigma_{2}\sigma_{1}\sigma_{2} \tag{7.40}%
\end{equation}
Accordingly,
\begin{equation}
a=\sigma_{1}\sigma_{2} \tag{7.41}%
\end{equation}
and,using Eq.(7.40), it is convenient to introduce yet another generator
\begin{equation}
b=a\sigma\tag{7.42}%
\end{equation}
Then,Eq.(7.39) is reduced to
\begin{equation}
a^{3}=b^{2} \tag{7.43}%
\end{equation}
Thus, we had obtained back the presentation group for the trefoil given
earlier by Eq.(7.3). $a^{n},$ introduced in Eq.(7.39), has some physical
meaning . If we would have a trivial braid made of n strands , then
application of $a^{n}$ \ to such trivial braid amounts of full rotation of the
lover side of the frame while keeping the upper side fixed[81]. Therefore , if
the system ,say, is \textbf{not} rotationally invariant, one cannot use the
concept of braids while the concept of fundamental group of free generators
still survives. The connections between braids and free groups could be made
more explicit if needed . Here we would like only to touch upon the most
immediate connections. To this purpose we need to introduce the notion of
\textbf{colored} braid. Basically, the braid is colored if the end of each
strand lies exactly over its beginning (that is for the ordered sequence on
the bottom frame there is exactly the same ordered sequence on the top). From
this definition, we notice that colored braids correspond to the
\textbf{distinguishable} particles in physics terminology . For the colored
braids P$_{n}$ made of n strands ,connection with free groups becomes very
direct. Fig. 25%
\begin{figure}
[ptb]
\begin{center}
\includegraphics[
height=2.2857in,
width=1.7867in
]%
{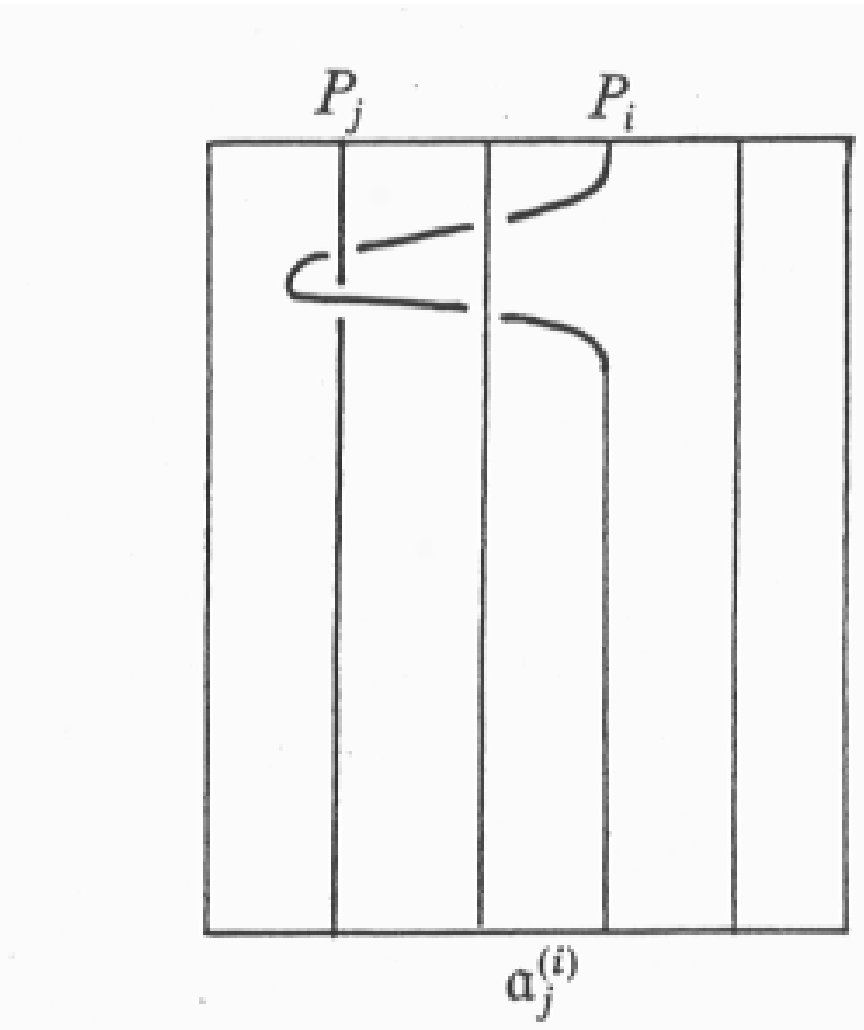}%
\caption{Action of the free group braiding generator a$_{j}^{(i)}$ on the
braid }%
\end{center}
\end{figure}
provides an example of a generic action of the braiding generator $a_{j}%
^{(i)}$ (an element of a \textbf{free} group) while Fig.26
\begin{figure}
[ptbptb]
\begin{center}
\includegraphics[
height=1.9623in,
width=4.2575in
]%
{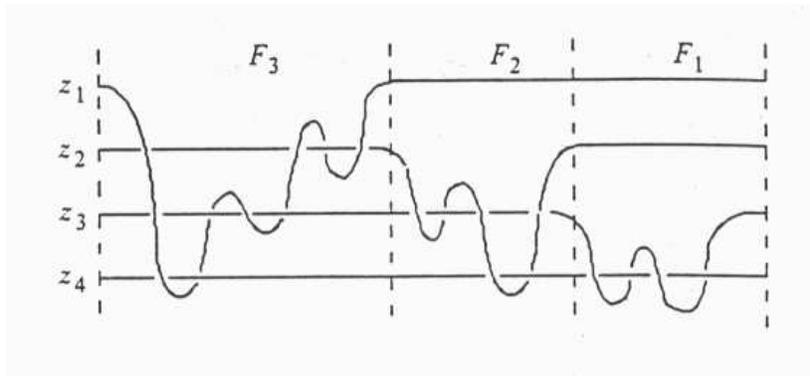}%
\caption{Colored braid made with help of composition of actions of generators
of free groups F$_{1},$F$_{2}$ and F$_{3}$ made of one , two and three
generators respectively}%
\end{center}
\end{figure}
illustrates how the colored braid P$_{4}$ could be made out of composition of
braids made with help of the generators of the free groups $F_{1},F_{2}$ and
$F_{3}.$ For general n mathematically,such braid can be written as semidirect
product
\begin{equation}
\text{P}_{n}=F_{1}\triangleleft(F_{2}\triangleleft(....(F_{n-1}\triangleleft
F_{n})) \tag{7.44}%
\end{equation}
For the colorless case, one can analyze the connections further by using Eq.s(7.36)-(7.39).

\pagebreak 

\bigskip

\ \ \ \ \ \ \ \ \ \ \ \ \ \ \ \ 

\bigskip\ \ \ \ \ \ \ \ \ \ \ \ \ \ \ \ \ \ \ \ \ \ \ \ \ \ \ \ \ \ \ \ \ \ \ \ \ \ \ \ \ \ \ \textbf{References}

[1] A.Kholodenko, Phys.Rev.E \textbf{58} ,R5213 (1998).

[2] F.Spitzer, Am.Math.Soc.Trans.\textbf{87}, 187 (1958).

[3] M.Coornaert and A.Papadopoulos, $Symbolic$ $Dynamics$ $and$ $Hyperbolic$

$Groups$ (Springer-Verlag,Berlin, 1993).

[4] R.Rammal,G.Tolouse and M.Virasoro, Rev.Mod.Phys.\textbf{58} ,765 (1986).

[5] D.Cartwright,V,Kaimanovich and W.Woess, Ann .Inst.Fourier(Grenoble)
\textbf{44},

1243(1994).

[6] R.Adler,Am.Math.Soc.Bull. \textbf{35},1 (1998).

[7] S.Wu Qian,Zhi-Y Gu and Guo-Q Xie, J.Phys.A\textbf{30} ,1273 (1997).

[8] Y.Aharonov and D.Bohm, Phys.Rev.\textbf{115}, 485 (1959).

[9] T.Shigehara,Phys.Rev.E \textbf{50},4357 (1994).

[10] T.Shigehara and T.Cheon, Phys.Rev. E \textbf{54}, 1321 (1996).

[11] T.Cheon and T.Shigevara ,Phys.Rev E \textbf{54} ,3300 (1996).

[12] R.Weaver and D.Sornette, Phys.Rev.E \textbf{52} ,3341 (1995).

[13] T.Aste, Phys.Rev.E \textbf{55},6233 (1997).

[14] A.Kholodenko and Th.Vilgis, Physics Reports \textbf{298} (5,6),251 (1998).

[15] N.Varopoulos, Math Proc. Cambr. Phil.Soc. \textbf{97},299 (1985).

[16] A.Katok and B.Hasselblat, \textit{Introduction to the Modern Theory of }

\textit{Dynamical Systems} (Cambridge U.Press,Cambridge,1997).

[17] A.Kholodenko , hep-th/9901040 and hep-th/9901047 , Journal of Geometry and

Physics(in press).

[18]K.Ito and H.McKean, \textit{Diffusion Processes and Their Sample Paths}

(Springer-Verlag, Berlin,1965).

[19] T.Lyons and H.McKean ,Adv. Math. \textbf{51},212 (1984).

[20] H.McKean and D.Sullivan ,Adv Math.\textbf{51}, 203 (1984).

[21] W.Woess, London Math.Soc.Bull. \textbf{26},1 (1994).

[22] L.Schnieps Editor,\textit{The Grotendieck Theory of Dessins D'Denfants }

(Cambridge U.Press, Cambridge, 1994).

[23] P.Kasteleyn, in Group Theoretical Methods in Physics

(Springer-Verlag, Berlin,1983).

[24] W.Thurston, Am.Math.Soc.Bull. \textbf{19},417 (1988).

[25] T.Cusick and M.Flahive, The Markoff and Lagrange Spectra (AMS

Publ.Providence, RI , 1989).

[26] L.Sculman, Phys.Rev.\textbf{176},1558 (1968).

[27] L.Schulman, J.Math.Phys. \textbf{12}, 304 (1971).

[28] J.Dovker,J.Phys. A \textbf{5} ,936 (1972).

[29] H.Poincare, \textit{Papers on Fuchsian Equations}, J.Stillwell

Editor(Springer-Verlag,Berlin,1985).

[30] M.Yoshida ,\textit{Fuchsian Differential Equations} (Vieweg \& Sohn,

Wiesbaden,1987).

[31] L.Schulman, \textit{Techniques and Applications of Path Integration}
(J.Wiley \&

Sons, New York,1981).

[32] R.Jackiw, Comments in Nuclear and Particle Phys.\textbf{15}, 99 (1985).

[33] P.Levy, \textit{Processes Stochasiques et Movement Brownien}

(Gauthier-Villars,Paris,1948).

[34] R.Durrett,\textit{ Brownian Motion and Martingales} (Wadsworth
Inc.,Belmont, CA,

1984).

[35] S.Deser,R.Jackiw and G.t'Hooft ,Ann.Phys.1\textbf{52}, 220 (1994).

[36] T.Jaroszkevich, Phys.Rev.Lett. \textbf{61}, 2401 (1988).

[37] D.Khandekar and S.Lavande ,Phys.Reports \textbf{137},115 (1986).

[38] H.Bateman and A.Erdelyi, \textit{Tables of Integral Transforms},Vol.1 (McGraw

Hill,New York,1954).

[39] N.Levinson and R.Redheffer, \textit{Complex Variables} (Holden Day Inc.,San

Francisco,1970).

[40] A.Kholodenko and A.Beyerlein, Phys.Lett.A \textbf{132 },347 (1988).

[41] P.Gerl, Lecture Notes in Math. Vol.1210 ,285 (1986).

[42] C.Itzykson and J.Drouffe, \textit{Statistical Field Theory}, Vol.1 (Cambridge

U.Press,Cambridge,1989).

[43] N.Gilbert and T.Porter , \textit{Knots and Surfaces} (Clarendon
Press,Oxford, 1994).

[44] W.Magnus, A.Karras and D.Solitar ,\textit{Combinatorial Group Theory}
(J.Wiley \&

Sons, New York,1966).

[45] H.Coxter and W.Moser ,\textit{Generators and Relations for Discrete}

\textit{Groups} (Springer-Verlag, Berlin,1980).

[46] S.Katok, \textit{Fuhsian Groups}(U.of Chicago Press, Chicago, 1992).

[47] L.Ford, \textit{Automorphic Functions} (McGraw Hill,New York,1929).

[48] Z.Nehari,\textit{Conformal Mapping} (Dower Publ.Inc.,New York,1975).

[49] L.Keen, Math.Intellegencer \textbf{16}, 11 (1994).

[50] J.Hempel,London Math.Soc.Bull.\textbf{20} ,97 (1986).

[51] N.Akhieser,\textit{ Elements of the Theory of Elliptic Functions} (Nauka,

Moscow,1970).

[52] P.Zograf and L.Takhtadjian, Math.USSR,Sbornik \textbf{60},143 (1988).

[53] A.Belavin,A.Polyakov and A.Zamolodchikov ,Nucl.Phys.B\textbf{ 241},333 (1984).

[54] L.Takhtadjian,Phys.Lett.A 8,3529 (1993).

[55] P.Zograpf,Leningrad Math.Journ.\textbf{1},941 (1990).

[56] C.Itzykson and J-B Zuber,\textit{Quantum Field Theory }(McGraw\ Hill,New

York,1980).

[57] L.Schlessinger,J.fur Reine und Angev.Math.\textbf{141},96 (1912).

[58] M.Jimbo and T.Miva , \textit{Integrable Systems in Statistical Mechanics
}(World

Scientific,Singapore,1985).

[59] B.Blok and S.Yankelowitz, Nucl.Phys. B \textbf{321},717 (1989).

[60] N.Hitchin, J.Diff.Geometry \textbf{42},30 (1995).

[61] V.Knizhnik and A.Zamolodchikov ,Nucl.Phys.B \textbf{247},83 (1984).

[62] C.Kassel,\textit{Quantum Groups} (Springer-Verlag,Berlin,1995).

[63] V.Drinfeld,Leningrad math.Journ.\textbf{1},1419 (1990).

[64] T.Lee and J.Murakami, Compositio Math.\textbf{102},41 (1996).

[65] D.Bar Natan,Topology \textbf{34}, 423 (1995).

[66] D.Altschuller and L.Freidel ,Comm.Math.Phys. \textbf{170}, 41 (1995).

[67] D.Bar Natan, Proc.Symp.in Appl.Math. \textbf{51},129 (1996).

[68] N.Muskhelishvili,\textit{Singular Integral equations} (Gr\"{o}ningen, Nederlands,1953).

[69] D.Hilbert, Gott.Nachr. \textbf{14},289(1900).

[70] G.Birkhoff,C\textit{ollected Works} ,Vol.1 (AMS Press,New York,1950).

[71]J.Plemej,\textit{Problems in the Sence of Riemann and Klein}(Interscience Publ.New

York,1964).

[72] S.Novikov,S.Manakov,L.Pitaevskii and V.Zakharov ,\textit{Theory of Solitons}

(Consultants Bureau, New York,1984).

[73] H.Kesten, Am.Math.Soc.Transactions \textbf{92},336 (1959).

[74] J.Stillwell, \textit{Classical Topology and Combinatorial Group}

\textit{Theory} (Springer-Verlag,Berlin,1995).

[75] J.Milnor, J.of Differential Geometry \textbf{2},1 (1968);see also
\textit{Collected Works}, Vol.1

(Publish or Perish ,Houston,1994).

[76] E.Gys,P.de la Harpe Editors ,\textit{Sur les Groupes Hyperboliques
D'Apres Mikhael}

\textit{Gromov}(Birkh\"{a}user,Boston ,1990 ).

[77] M.Coornaert,T.Deslant and A.Papadopoulos ,\textit{Geometrie et Theorie des}

\textit{Groups }(Springer-Verlag,Berlin,1990).

[78] W.Feller, \textit{An Introduction to Probability Theory and its
Applications },Vol.1

(John Wiley \& Sons Inc.,New York,1957).

[79] B.Huges and M.Sahimi , J.Stat.Phys.\textbf{29},781 (1982).

[80] N.Biggs,\textit{ Iteraction Models} (Cambridge U.Press,Cambridge,1977).

[81] G,Brude and H.Zieshang , \textit{Knots} (Walter de Gruyter,Berlin,1985) .

[82] W.Magnus, \textit{Max Dehn} ,Math.Intellegencer,\textbf{1},132 (1978).

[83] T.Grossman and W.Magnus ,\textit{Groups and Their Graphs} (Random House,New

York,1964).

[84] W.Thurston,Am.Math.Soc.Bulletin \textbf{6},357 (1982).

[85] \ M.Gutzviller,\textit{Chaos in Classical and Quantum Mechanics}

(Springer-Verlag,Berlin,1990).

[86] J.Avron,R.Seiler and P.Zograf,Phys.Rev.Lett.\textbf{75},697 (1995).

[87] L.Gullope and M.Zworski, Ann.Math.\textbf{145},597 (1997); ibid Y.Minsky,\textit{The}

\textit{Classification of Punctured Torus Groups} ,Ann.Math. (to be published).

[88] C.Series,Math.Intellegencer \textbf{7}, 20 (1985).

[89] B.Kitchens, Symbolic Dynamics (Springer-Verlag,Berlin,1988).

[90] C.Series, J.London Math.Soc. \textbf{31} ,69 (1985).

[91] G.Hardy and E.Wright, \textit{Introduction to the Theory of Numbers }(Clarendon

Press,Oxford,1960).

[92] S.Nonnemacher and A.Voros, J.Stat.Phys.\textbf{ 92},431 (1998).

[93] P.di Francesco, P.Mathieu and D.Senechal, \textit{Conformal Field}

\textit{Theory }(Springer-Verlag,Berlin,1997).

[94] J.Marklof,Comm.Math.Phys.\textbf{199},169 (1998).

[95] J.Milnor, \textit{Singular Points of Complex Hypersurfaces }(Princeton

U.Press,Princeton,1968).

[96] W.Abikoff, \textit{The Real Analytic Theory of Teichm\"{u}ller Space}

(Sringer-Verlag,Berlin,1980).

[97] J.Birman ,in \textit{Discontinuous Groups and Riemann Surfaces}(Princeton

U.Press,Princeton,1974).

[98] A.Casson and S.Bleiler,\textit{Automorphisms of Surfaces after Nielsen and}

\textit{Thurston}(Cambridge U.Press,Cambridge,1995).

[99] P.Chaikin and D.Lubensky \textit{,Principles of Condensed Matter Physics} (Cambridge

U.Press,Cambridge,1995).

[100] C.McMullen , \textit{Renormalization and 3-Manifolds Which Fiber Over the}

\textit{Circle}(Princeton U.Press,Princeton,1996).

[101] Y.Pesin, \textit{Dimension Theory in Dynamical Systems }(The U.of Chicago

Press,Chicago,1997).

[102] R.Ghrist,P.Holmes and M.Sullivan, \textit{Knots, Links and Three-Dimensional}

\textit{Flows}(Springer-Verlag,Berlin,1997).

[103] W.Thurston,\textit{Geometry and Topology of 3 Manifolds} (Princeton U.,Department

of Mathematics,Set of Lecture Notes,Princeton,1979).

[104] D.Rolfsen,\textit{Knots and Links} (Publish or Perish,Houston,TX,1990).

[105] A.Kholodenko, hep-th/9902167.

[106] P.Lax and R.Phillips, \textit{Scattering Theory for Automorphic
Functions} (Princeton

U.Press,Princeton,1976).

[107] H.Farkas and I.Kra , \textit{Riemann Surfaces }(Springer-Verlag,New York,1992).

[108] R.Fricke,G\"{o}tt.Nach.12 ,91 (1896).

[109] F,Wilzek,\textit{Fractional Statistics and Anyon Superconductivity }(World

Scientific,Singapore,1990).

[110] H.Kohn, in Lecture Notes in Pure and Applied Math,Vol.92 ,D.Chudnovsky and

G.Chudnovsky Editoirs (Marcell Dekker,New York,1984).

[111] B.Bowdich,London Math.Soc.Bull.\textbf{28},73 (1996).

[112] B.Bowdich, London Math Soc.Proceedings,\textbf{77},697 (1998).

[113] A.Hurvitz, Math.Ann.39,279 (1891).

[114] F.Hirzerbruch and D.Zagier,\textit{The Atiach-Singer Theorem and Elementary}

\textit{Number} \textit{Theory}(Publish or Perish,Houston,1974).

[115] G.Belyi, Math.USSR, Izvestia \textbf{14},247 (1980).

[116] P.Cohen,C.Itzykson and J.Wolfart,Comm.Math.Phys.\textbf{163} ,605 (1994).

[117] P.Richens and M.Berry ,Physica D \textbf{2} ,495 (1981).

[118] A.Zemlyakov and A.Katok ,Math.Notes,\textbf{18},291 (1975).

[119] S.Tabachnikov,\textit{ Billiards }(SMF,Marseille,1995).

[120] G.Shabat and V.Voevodsky, in Progress in Math.,Vol.88

(Birkh\"{a}user,Boston,1990).

[121] E.Artin,Ann.Math.\textbf{48},101;ibid 643 (1947).
\end{document}